%% file: main.tex
\documentclass[twocolumn]{aastex631}
\usepackage{}

\usepackage{gensymb}  

\usepackage{multirow}  

\submitjournal{ApJ}

\shorttitle{JWST Spectra of MACS0647--JD}
\shortauthors{Hsiao et al.}

\begin{document}

\title{\JWST\ NIRSpec spectroscopy of the triply-lensed $z = 10.17$ galaxy MACS0647--JD}

\correspondingauthor{Tiger Hsiao and Abdurro'uf}
\email{tiger.hsiao@cfa.harvard.edu, fabdurr1@jhu.edu}


\newcommand{\STScI}{\affiliation{Space Telescope Science Institute (STScI), 3700 San Martin Drive, Baltimore, MD 21218, USA}}

\newcommand{\JHU}{\affiliation{Center for Astrophysical Sciences, Department of Physics and Astronomy, The Johns Hopkins University, 3400 N Charles St. Baltimore, MD 21218, USA}}

\newcommand{\ESAAURA}{\affiliation{Association of Universities for Research in Astronomy (AURA), Inc.~for the European Space Agency (ESA)}}

\newcommand{\UTAustin}{\affiliation{The University of Texas at Austin, Department of Astronomy, Austin, TX, United States}}

\newcommand{\DAWN}{\affiliation{Cosmic Dawn Center (DAWN), Copenhagen, Denmark}}

\newcommand{\NielsBohr}{\affiliation{Niels Bohr Institute, University of Copenhagen, Jagtvej 128, Copenhagen, Denmark}}

\newcommand{\Groningen}{\affiliation{Kapteyn Astronomical Institute, University of Groningen, P.O. Box 800, 9700AV Groningen, The Netherlands}}

\newcommand{\NASAGoddard}{\affiliation{Observational Cosmology Lab, NASA Goddard Space Flight Center, Greenbelt, MD 20771, USA}}

\newcommand{\Uppsala}{\affiliation{Observational Astrophysics, Department of Physics and Astronomy, Uppsala University, Box 516, SE-751 20 Uppsala, Sweden}}

\newcommand{\INAFOAS}{\affiliation{INAF -- OAS, Osservatorio di Astrofisica e Scienza dello Spazio di Bologna, via Gobetti 93/3, I-40129 Bologna, Italy}}

\newcommand{\UMD}{\affiliation{Department of Astronomy, University of Maryland, College Park, 20742, USA}}

\newcommand{\CSIC}{\affiliation{Instituto de F\'isica de Cantabria (CSIC-UC). Avda. Los Castros s/n. 39005 Santander, Spain}}

\newcommand{\Sussex}{\affiliation{Astronomy Centre, University of Sussex, Falmer, Brighton BN1 9QH, UK}}

\newcommand{\Malta}{\affiliation{Institute of Space Sciences and Astronomy, University of Malta, Msida MSD 2080, Malta}}

\newcommand{\Ljubljana}{\affiliation{Department of Mathematics and Physics, University of Ljubljana, Jadranska ulica 19, SI-1000 Ljubljana, Slovenia}}
\newcommand{\UCDavis}{\affiliation{Department of Physics and Astronomy, University of California, Davis, 1 Shields Ave, Davis, CA 95616, USA}}
\newcommand{\LBNL}{\affiliation{Lawrence Berkeley National Laboratory, CA 94720, USA}}
\newcommand{\Auckland}{\affiliation{Department of Physics, University of Auckland, Private Bag 92019, Auckland, New Zealand}}
\newcommand{\Manchester}{\affiliation{Jodrell Bank Centre for Astrophysics, University of Manchester, Oxford Road, Manchester M13 9PL, UK}}
\newcommand{\Edinburgh}{\affiliation{Institute for Astronomy, University of Edinburgh, Royal Observatory, Edinburgh EH9 3HJ, UK}}
\newcommand{\BenGurion}{\affiliation{Physics Department, Ben-Gurion University of the Negev, P.O. Box 653, Be'er-Sheva 84105, Israel}}
\newcommand{\TAMUG}{\affiliation{George P. and Cynthia Woods Mitchell Institute for Fundamental Physics and Astronomy, Texas A\&M University, College Station, TX 78743, USA}}
\newcommand{\TAMU}{\affiliation{Department of Physics and Astronomy, Texas A\&M University, 4242 TAMU, College Station, TX 78743, USA}}
\newcommand{\BCM}{\affiliation{Department of Liberal Arts and Sciences, Berklee College of Music, 7 Haviland Street, Boston, MA 02215, USA}}
\newcommand{\CfA}{\affiliation{Center for Astrophysics \text{\textbar} Harvard \& Smithsonian, 60 Garden Street, Cambridge, MA 02138, USA}}
\newcommand{\Stockholm}{\affiliation{Department of Astronomy, Oskar Klein Centre, Stockholm University, AlbaNova University Centre, SE-106 91 Stockholm, Sweden}}
\newcommand{\MichiganState}{\affiliation{Michigan State University, Physics \& Astronomy Department, East Lansing, MI, USA}}
\newcommand{\Basque}{\affiliation{Department of Theoretical Physics, University of the Basque Country UPV-EHU, E-48040 Bilbao, Spain}}
\newcommand{\DIPC}{\affiliation{Donostia International Physics Center (DIPC), 20018 Donostia, The Basque Country, Spain}}
\newcommand{\IKERBASQUE}{\affiliation{IKERBASQUE, Basque Foundation for Science, Alameda Urquijo, 36-5 E-48008 Bilbao, Spain}}
\newcommand{\ASU}{\affiliation{School of Earth and Space Exploration, Arizona State University, Tempe, AZ 85287-6004, USA}}
\newcommand{\Northwestern}{\affiliation{Department of Physics and Astronomy, Northwestern University, 2145 Sheridan Road, Evanston, IL, 60208, USA}}
\newcommand{\RIT}{\affiliation{School of Physics and Astronomy, Rochester Institute of Technology, 84 Lomb Memorial Drive, Rochester, NY 14623, USA}}
\newcommand{\CIERA}{\affiliation{Center for Interdisciplinary Exploration and Research in Astrophysics (CIERA), Northwestern University, 1800 Sherman Avenue, Evanston, IL, 60201, USA.}}
\newcommand{\Clay}{\affiliation{Clay Center Observatory, Dexter Southfield, 20 Newton Street, Brookline, MA 02445, USA}}

\newcommand{\NSFFellow}{\altaffiliation{NSF Graduate Fellow}}
\newcommand{\NPPFellow}{\altaffiliation{NASA Postdoctoral Fellow}}
\newcommand{\HubbleFellow}{\altaffiliation{Hubble Fellow}}
\newcommand{\EqualContributions}{\altaffiliation{Both lead authors contributed equally}}


\author[0000-0003-4512-8705]{Tiger Yu-Yang Hsiao} \EqualContributions \CfA \JHU \STScI 
\author[0000-0002-5258-8761]{Abdurro'uf} \EqualContributions \JHU \STScI
\author[0000-0001-7410-7669]{Dan Coe} \STScI \ESAAURA \JHU
\author[0000-0003-2366-8858]{Rebecca L. Larson} \RIT
\author[0000-0003-1187-4240]{Intae Jung} \STScI
\author[0000-0003-2589-762X]{Matilde Mingozzi} \STScI
\author[0000-0001-8460-1564]{Pratika Dayal} \Groningen
\author[0000-0002-5320-2568]{Nimisha Kumari} \STScI \ESAAURA
\author[0000-0002-5588-9156]{Vasily Kokorev} \UTAustin
\author[0000-0002-4853-1076]{Anton Vikaeus} \Uppsala
\author[0000-0003-2680-005X]{Gabriel Brammer} \DAWN \NielsBohr
\author[0000-0001-6278-032X]{Lukas J. Furtak} \BenGurion

\author[0000-0002-8192-8091]{Angela Adamo} \Stockholm

\author[0000-0002-8144-9285]{Felipe Andrade-Santos} \BCM \CfA \Clay
\author[0000-0002-0243-6575]{Jacqueline Antwi-Danso} \TAMUG \TAMU
\author[0000-0001-5984-0395]{Maru\v{s}a Brada\v{c}} \Ljubljana \UCDavis
\author[0000-0002-7908-9284]{Larry D. Bradley} \STScI
\author[0000-0002-8785-8979]{Tom Broadhurst} \Basque \DIPC \IKERBASQUE
\author[0000-0002-1482-5818]{Adam C. Carnall} \Edinburgh
\author[0000-0003-1949-7638]{Christopher J. Conselice} \Manchester
\author[0000-0001-9065-3926]{Jose M. Diego} \CSIC
\author[0000-0002-2808-0853]{Megan Donahue} \MichiganState
\author[0000-0002-1722-6343]{Jan J. Eldridge} \Auckland
\author[0000-0001-7201-5066]{Seiji Fujimoto}\HubbleFellow \UTAustin
\author[0000-0002-6586-4446]{Alaina Henry} \STScI \JHU
\author[0000-0003-4857-8699]{Svea Hernandez} \ESAAURA \STScI
\author[0000-0001-6251-4988]{Taylor A. Hutchison} \NPPFellow \NASAGoddard
\author[0000-0003-4372-2006]{Bethan L. James} \STScI \ESAAURA
\author[0000-0002-5222-5717]{Colin Norman} \JHU \STScI
\author[0000-0002-7464-7857]{Hyunbae Park} \LBNL
\author[0000-0003-3382-5941]{Norbert Pirzkal} \STScI
\author[0000-0002-9365-7989]{Marc Postman} \STScI
\author[0000-0003-4223-7324]{Massimo Ricotti} \UMD
\author[0000-0002-7627-6551]{Jane R.~Rigby} \NASAGoddard
\author[0000-0002-5057-135X]{Eros Vanzella} \INAFOAS
\author[0000-0003-1815-0114]{Brian Welch} \NASAGoddard \UMD
\author[0000-0003-3903-6935]{Stephen M.~Wilkins} \Sussex \Malta
\author[0000-0001-8156-6281]{Rogier A. Windhorst} \ASU
\author[0000-0002-9217-7051]{Xinfeng Xu} \Northwestern \CIERA
\author[0000-0003-1096-2636]{Erik Zackrisson} \Uppsala
\author[0000-0002-0350-4488]{Adi Zitrin} \BenGurion 

\input{newcommands}

\begin{abstract}

We present \JWST/NIRSpec prism spectroscopy of \JD,
the triply-lensed $z \sim 11$ candidate discovered in \HST\ imaging
and spatially resolved by \JWST\ imaging into two components A and B.
%
Spectroscopy of component A yields
a spectroscopic redshift $z=10.17$ 
based on 7 detected emission lines:
\CIIIdw, \OIIw, \NeIIIw, \NeIIIwb, \Hdeltaw, \Hgammaw, and \OIIIwa.
These are the second-most distant detections of these emission lines to date, 
in a galaxy observed just 460 million years after the Big Bang. 
%
Based on observed and extrapolated line flux ratios
we derive a
gas-phase metallicity \logOH\ $\sim$ 7.5 -- 8.0, or $Z\sim$~(0.06 -- 0.2) \Zsun,
ionization parameter \logU\  $ = -1.9\pm0.2$,
%
and an ionizing photon production efficiency 
${\rm log}(\xi_{\rm ion})=25.2\pm0.2\,$erg\inv\ Hz.
The spectrum has a softened \Lya\ break,
evidence for a strong {\Lya} damping wing.
The {\Lya} damping wing also suppresses the F150W photometry,
explaining the slightly overestimated photometric redshift $z = 10.6 \pm 0.3$.
%
\JD\ has a stellar mass log($M/M_\odot$) = $8.1 \pm 0.3$,
including \about\ 6$\times$\tentothe{7} \Msun\ in component A,
most of which formed recently (within \about 20 Myr)
with a star formation rate SFR$\sim 2\pm1$ \Msun\ yr\inv, all
within an effective radius $70\pm24\,$pc.
%
%
%
Spectroscopy of a fainter companion galaxy C separated by a distance of \about\ 3$\,$kpc 
reveals a Lyman break consistent with $z \sim 10.17$.
\JD\ is likely the most distant galaxy merger known.

\end{abstract}


\section{Introduction}
\label{sec:intro}
Spectroscopy using the NIRSpec instrument \citep{NIRSpec_Jakobsen2022,NIRSpec-MOS_Ferruit2022,Boker2023} onboard \JWST\ \citep{Gardner_23PASP, Rigby_23PASP} 
is beginning to confirm distant galaxy candidates
and reveal their properties in the early universe
\citep[$z>8$; e.g.,][]{Fujimoto2023,Roberts2022, Bunker2023, Arrabal-Haro2023a,Harikane2023,CurtisLake2023}.
With coverage out to $5.3\,$\um, NIRSpec can detect the strong emission lines \OIIIww\ and \Hbeta\ in galaxies as distant as $z \sim 9.5$,
confirming redshifts and constraining properties 
such as metallicity and ionization parameter 
when combined with other emission lines
\citepeg{Williams2022,Nakajima2023,Boyett2023,Cameron2023a,Tang2023,Sanders2023,Jung2023}.

At $z > 10$, within the first 500 million years,
spectroscopic confirmations and detailed studies are more challenging because \OIII +\Hbeta\ is redshifted out of the NIRSpec wavelength range,
and fainter emission lines are difficult to detect in low-luminosity $z > 10$ galaxies.
Brighter galaxies or longer exposure times are required.
\cite{CurtisLake2023} presented deep spectra with a spectral resolution of $R\sim100$ that impressively confirmed galaxies with redshifts $z = 10.4$, 11.6, 12.6, and 13.2 based on significant Lyman break detections  and an absence of definitive emission lines in long exposures of 9 -- 28 hours for these faint candidates ({F200W} AB mag 28 -- 29).
Similarly, \cite{Arrabal-Haro2023a} confirmed galaxies
at $z = 10.1$ and $z = 11.4$ based on Lyman break detections
with no emission lines in 5-hour NIRSpec prism spectra
that also revised the redshift of a $z \sim 16$ candidate down to $z = 4.912$.

Over a decade of Hubble Space Telescope (\HST) WFC3/IR observations 
yielded two $z > 10$ galaxies bright enough ({F200W} AB mag {$25.8\pm0.1$}) for more detailed study.
GN-z11 was one of the most distant galaxies ever discovered in \HST\ imaging \citep{Oesch16}.
They noted that given the search area, it was surprising to discover a galaxy so bright.
\JWST\ NIRCam \citep{Rieke2023} imaging photometry yields F200W AB mag of {$25.9\pm0.1$} and $M_{UV} = -21.6$,
or $M_{UV} = -21.8$ corrected for dust \citep{Tacchella2023}. 
\JWST\ NIRSpec spectroscopy has now confirmed GN-z11 at $z = 10.603$
and revealed 12 emission lines \citep{Bunker2023}.

\JD\ was the other similarly bright $z \sim 11$ candidate discovered in \HST\ imaging \citep{Coe2013},
supported by \HST\ grism observations \citep{Pirzkal2015}, \Spitzer\ imaging \citep{Lam2019},
and gravitational lens modeling \citep{Chan2017}.
MACS0647--JD is triply-lensed by a massive foreground galaxy cluster, 
MACSJ0647.7+7015 \citep[$z=0.591$;][]{Ebeling2007}.
Despite lensing magnifications of \about 8, 5, and 2 \citep{Meena2023},
the three lensed images JD1, JD2, and JD3 were spatially unresolved in \HST\ imaging.

\JWST/NIRCam imaging resolved \JD\ as having two small components 
in a possible galaxy merger at $z=10.6\pm0.3$ \citep{Hsiao2023}.
The brighter component A has an effective radius $r\sim70\pm24\,$pc 
and appears bluer,
{($\beta\sim-2.6\pm0.1$)}, 
likely due to its young stellar population (\about 50$\,$Myr old) and no dust.
The other component, JDB, has a smaller radius of $r\sim20^{+8}_{-5}\,$pc 
and appears redder {($\beta\sim-2\pm0.2$)}, 
, likely due to an older stellar population (\about 100$\,$Myr old) 
and mild dust ($A_{V}\sim 0.1\,$mag).
The different inferred star formation histories 
suggest these two components formed separately and are in the process of merging.
A third triply-lensed component C was also identified 2\arcsec\ away
(\about 3$\,$kpc in the delensed source plane).
With similar colors and a consistent photometric redshift, 
C may be destined to merge with A and B.

Compared to GN-z11, \JD\ is intrinsically fainter ($M_{UV} = -20.3$),
yet observed to be brighter; the brightest lensed image JD1 is F200W AB mag {$25.0\pm0.1$}.
The brightest clump A is $M_{UV} = -19.5$ and lensed to F200W AB mag of {$25.8\pm0.1$} in JD1.
This enables detailed study with \JWST.



In this paper, we report NIRSpec and NIRCam observations of MACS0647--JD
obtained by \JWST\ observation program GO 1433 (PI Coe).
Most of the NIRCam images were analyzed previously by \citet{Hsiao2023}.
This work includes a new F480M image, deeper F200W data, and updated image reductions and photometry
(\S\ref{sec:data}).
Based on NIRSpec prism ($R \sim 100$) spectroscopy spanning 0.7 -- 5.3$\,$\um,
we detect nebular emission lines,
measure the spectroscopic redshift,
and derive physical properties including the ionization parameter and metallicity 
using various emission line ratios 
as well as photometric spectral energy distribution (SED) fitting (\S\ref{sec:results&discussion}).
We present conclusions in \S\ref{sec:conclusions}.

\section{Observations}
\label{sec:data}
\JWST\ program GO 1433 (PI Coe) observed MACS0647$-$JD as summarized in Table \ref{tab:obs}.
We also analyze archival \HST\ imaging from programs GO 9722, 10493, 10793, 12101, and 13317,
described in \cite{Hsiao2023}.
All \JWST\ and \HST\ data are publicly available in the 
Mikulski Archive for Space Telescopes (MAST{; \dataset[DOI:10.17909/wpys-ap03]{https://archive.stsci.edu/doi/resolve/resolve.html?doi=10.17909/wpys-ap03}}).
We also provide reduced data products\footnote{\url{https://cosmic-spring.github.io}}
along with our data analysis scripts.\footnote{\url{https://github.com/cosmic-spring/MACS0647-JD-NIRSpec}}

\subsection{NIRSpec MOS prism spectroscopy}

There were two separate NIRSpec 
multi-object spectroscopy (MOS) observations (Obs 21 and Obs 23) 
using the microshutter assembly (MSA).
Both used the low-resolution $R \sim 30-300$ prism yielding data from 0.6 -- 5.3$\,$\um.
Obs 23 was performed with standard 3-slitlet nods
used to subtract backgrounds measured near each target.
The total exposure time was 1.8 hours.
Obs 21 was performed with single slitlets,
requiring backgrounds to be measured in nearby slitlets
designed to observe ``blank'' sky and intracluster light.
Two MSA configurations were executed in Obs 21
with a total exposure time of 1.8 hours.
The two configurations were similar with small dithers between them
and most targets included in both.

The two brightest lensed images of \JD\ (JD1 and JD2)
were observed in all exposures of all observations
for a total exposure time of 3.6 hours.
The slitlets primarily captured light from the brightest clump A.
The brightest lensed images of the companion galaxy C (JD1C and JD2C)
were also observed in all exposures.
Since JD1 and JD2 are multiple images of the same object 
with JD2 magnified \about 70\% as bright ($\mu\sim5.3$) as JD1,
the resulting {summed} spectra have signal-to-noise ratio (SNR) similar to 6-hour total effective exposure times on JD1A and JD1C.

Clump B was observed only in {Obs} 21 of JD2. Given its proximity to clump A, 
it is difficult to spatially resolve its contribution to the JD2 spectrum.
They are separated by only one or two 0.1\arcsec\ pixels in the cross-dispersion direction {(i.e., in the 2D NIRSpec spectrum)}.
We do perhaps detect some emission line {fluxes (i.e.,\CIIIdw)} from JD2B as we discuss in \S\ref{sec:CIII} {(see also the second row of Figure \ref{fig:spectra}).}

We retrieved NIRSpec Level 1 data products from MAST
and processed them with the STScI \JWST\ pipeline\footnoteurl{https://github.com/spacetelescope/jwst} version 1.9.2
and \msaexp\footnoteurl{https://github.com/gbrammer/msaexp} version 0.6.0.
First, \msaexp\ corrects for 1/$f$ noise\footnote{While the observations were obtained in IRS2 readout mode to mitigate the effect of 1/$f$ noise, NIRSpec data still exhibit this characteristic to some extent, motivating the correction.}, 
masks snowballs\footnote{Snowballs are now masked by the \JWST\ Level 1 pipeline as well.}, 
corrects bias levels in individual exposures, and rescales the noise array \tt{RNOISE} based on measurements in blank regions.
Next, we run \JWST\ pipeline Stage 2 routines to 
perform WCS registration, flat-fielding, pathloss corrections, and flux calibration.

For the 3-slitlet data, background subtraction is performed locally
before drizzling the 3 exposures onto a common 2D pixel grid.
For the single-slitlet data, we drizzle-combined data from the 2 exposures (MSA configurations).
Then we subtracted the 2D background spectrum in a nearby slit (\#50981) that observed a relatively blank region of the image.

To extract each 1D spectrum,
\msaexp\ uses a Gaussian profile in the cross-dispersion direction,
derived from an inverse-weighted sum of the 2D spectrum in the dispersion direction, 
based on the optimal extraction described in \cite{Horne1986}.
The results are shown in Figure \ref{fig:spectra} for each observation.
{The plots show the numbers for the extraction Gaussian: offset $\pm$ width in pixels.
There are no design choices to exclude any contribution from clumps A and B.}

Finally, we {sum} the 1D spectra, adding flux uncertainties in quadrature,
yielding the spectrum shown in Figure \ref{fig:spectrum}.
This spectrum includes the total flux observed in all 4 exposures for JD1A and JD2A.




\subsection{NIRCam imaging}

NIRCam imaging was obtained first in 6 filters, F115W, F150W, F200W, F277W, F365W, and F444W spanning 1--5$\,\mu$m, and the results were studied and presented in \citet{Hsiao2023}.
Additional NIRCam imaging has since been obtained in F200W and F480M,
concurrent with the NIRSpec {Obs} 21.
The F480M observation was included to observe \JD\ redward of the Balmer break
to better constrain its stellar age and mass.
Each NIRCam image has an exposure time of 35 minutes, 
split between 4 INTRAMODULEBOX dithers that cover the short-wavelength gaps.

We process the NIRCam images using the STScI \JWST\ pipeline and \grizli\ \citep{grizli}.
In  \citet{Hsiao2023}, we analyzed {the fourth release (v4) of the images reduced with \grizli} and then performed zeropoint corrections based on updated calibrations
that remain robust today \citep{Boyer2022}.
Those updated zeropoints were implemented in \grizli\ v5.
In this work, we analyze \grizli\ v6 image mosaics\footnote{\url{https://github.com/gbrammer/grizli/blob/master/docs/grizli/image-release-v6.rst}}
featuring improved sky flats, bad pixel tables, wisp templates, and long-wavelength zeropoints.

Briefly, in each NIRCam exposure, 
the \grizli\ pipeline applies corrections for $1/f$ noise striping
and masks ``snowballs''\footnote{\url{https://jwst-docs.stsci.edu/data-artifacts-and-features/snowballs-and-shower-artifacts}} 
and ``wisps''\footnote{\url{https://jwst-docs.stsci.edu/jwst-near-infrared-camera/nircam-instrument-features-and-caveats/nircam-claws-and-wisps}}.
Then \grizli\ drizzle-combines all exposures in each filter
on to a common pixel grid using \astrodrizzle\ \citep{MultiDrizzle,DrizzlePac}.
The NIRCam short wavelength images are drizzled to 0$\farcs$02 pixels,
and all other images are drizzled to 0$\farcs$04 pixels.
All images are aligned with coordinates registered to GAIA DR3 \citep{Gaia_EDR3}.
We create color images using \trilogy\footnote{\url{https://github.com/dancoe/trilogy}}
\citep{Coe12_Trilogy}.

\begin{figure*}
\includegraphics[width=2in]{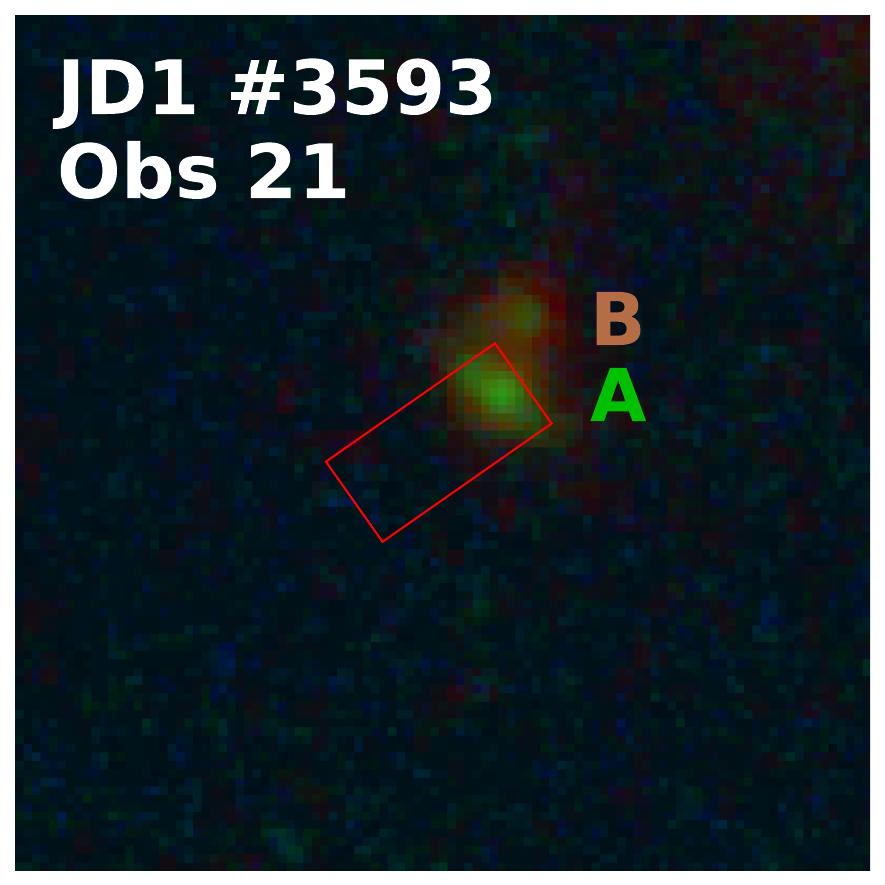}
\includegraphics[width=5in]{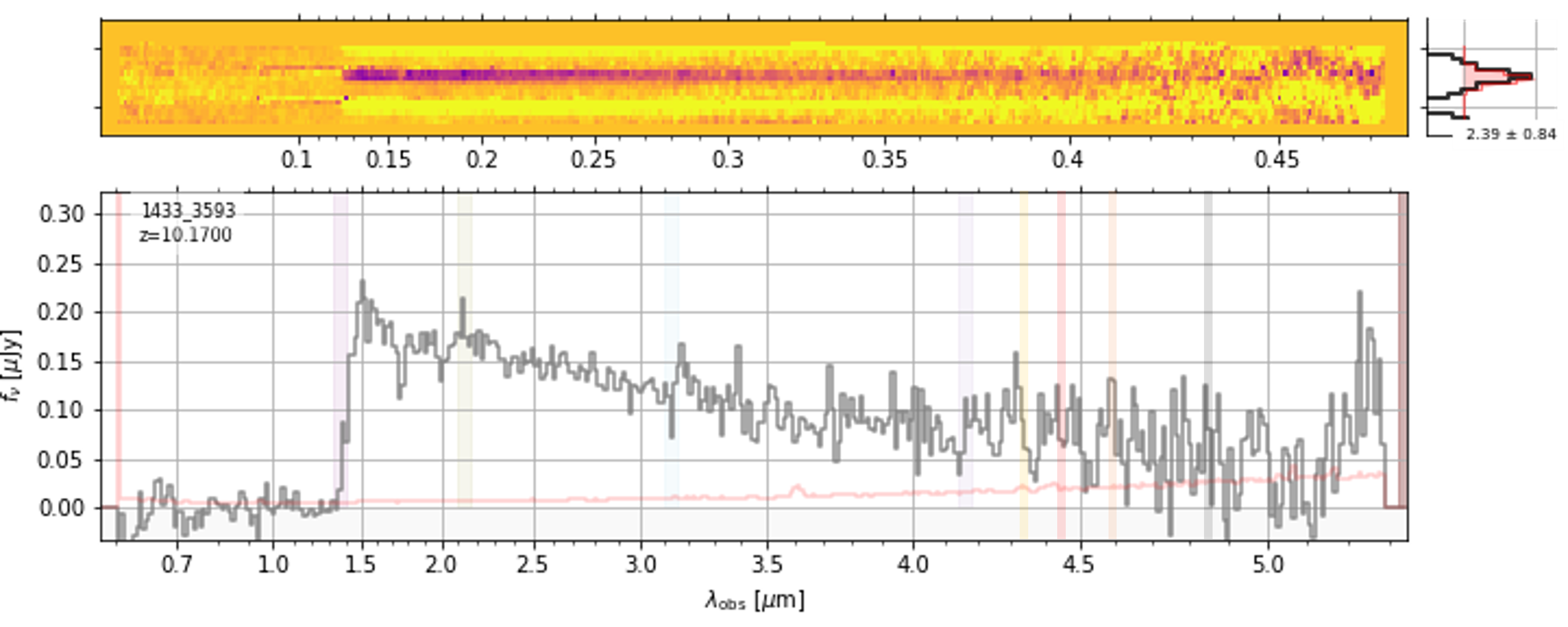}
\includegraphics[width=2in]{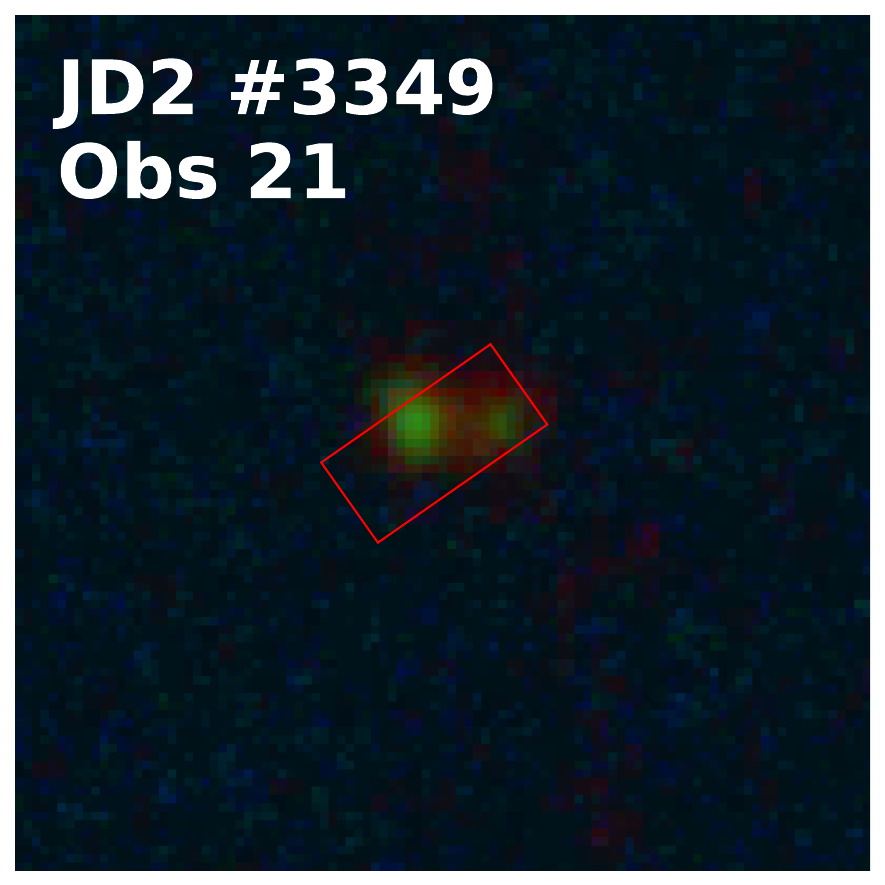}
\includegraphics[width=5in]{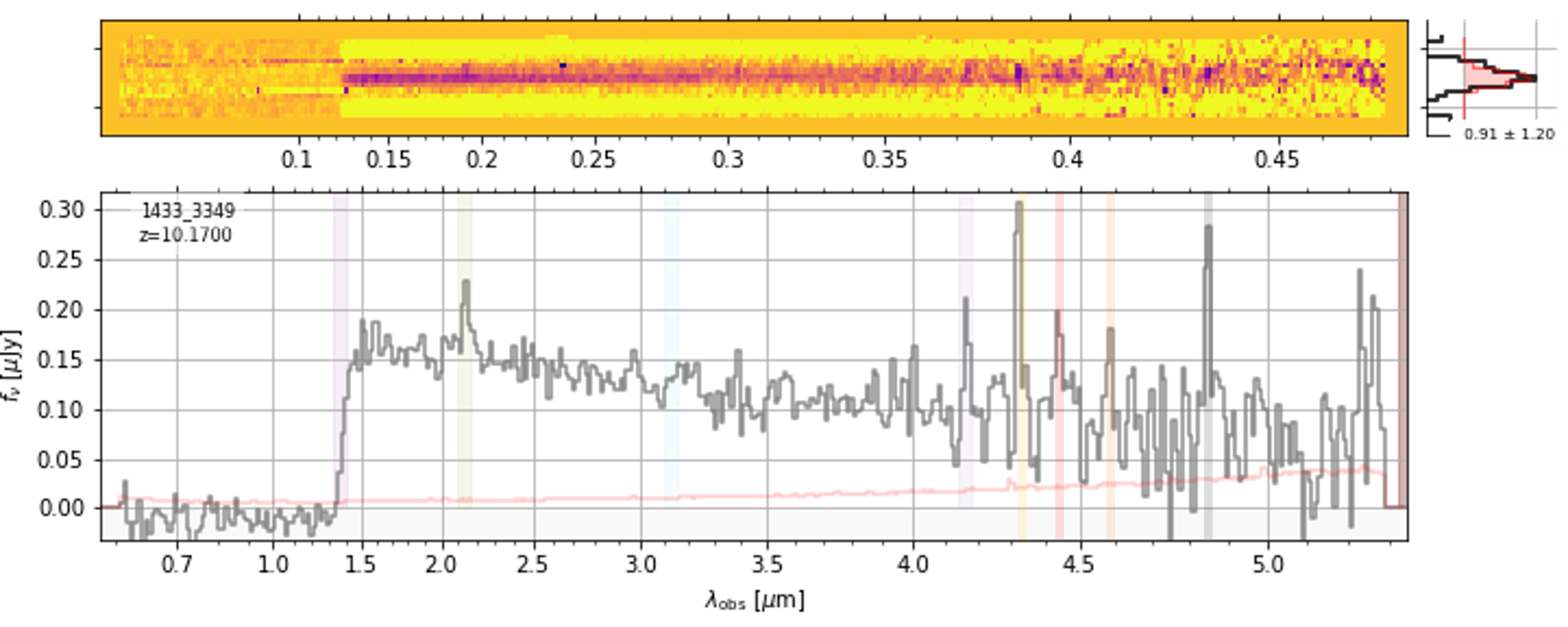}
\includegraphics[width=2in]{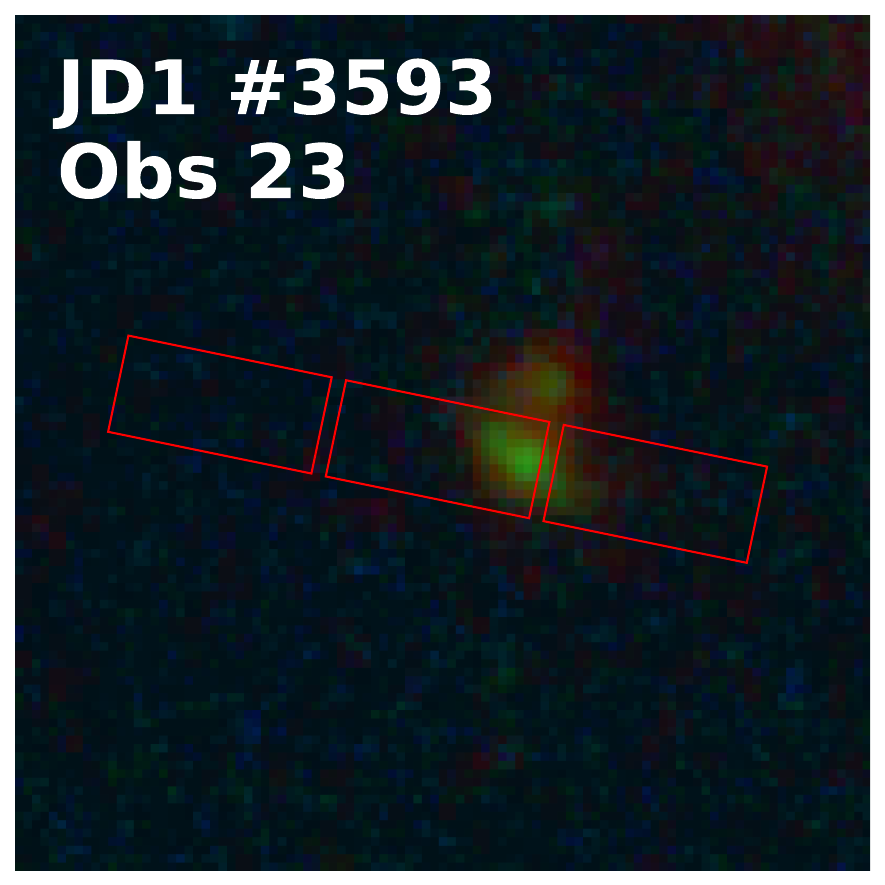}
\includegraphics[width=5in]{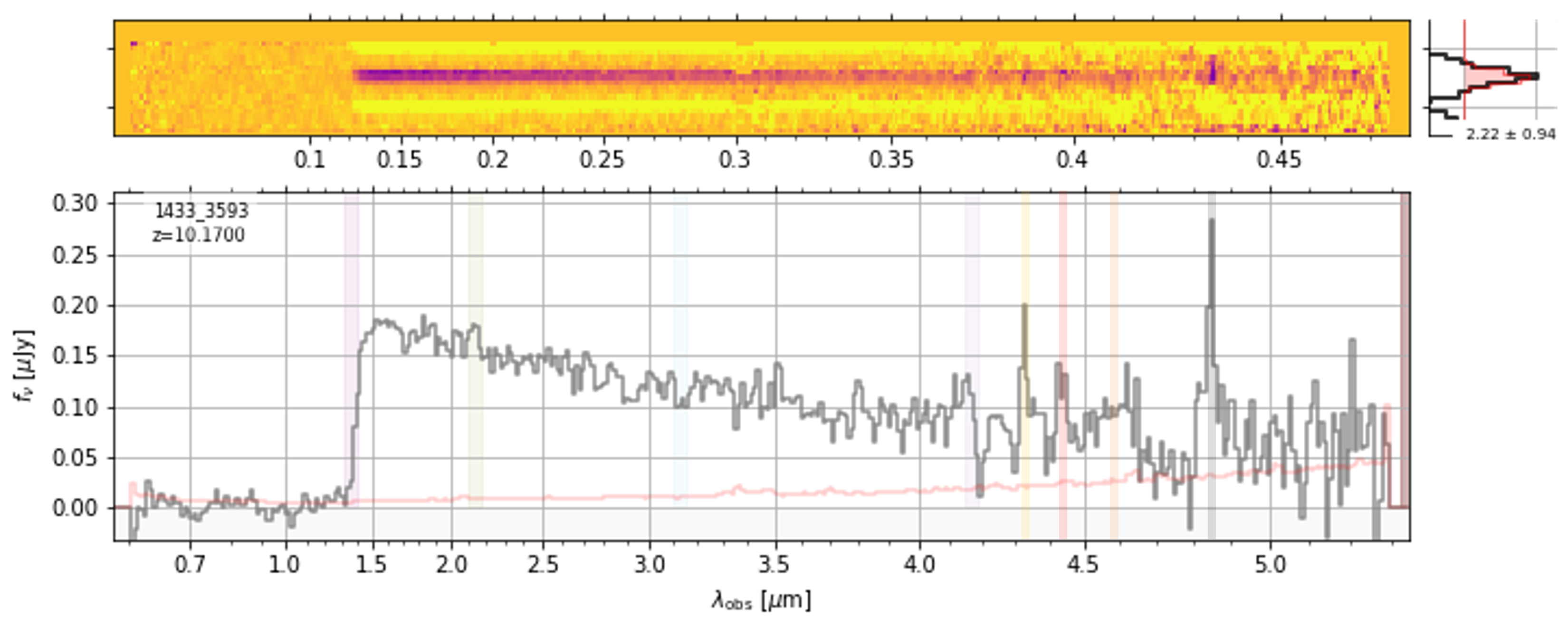}
\includegraphics[width=2in]{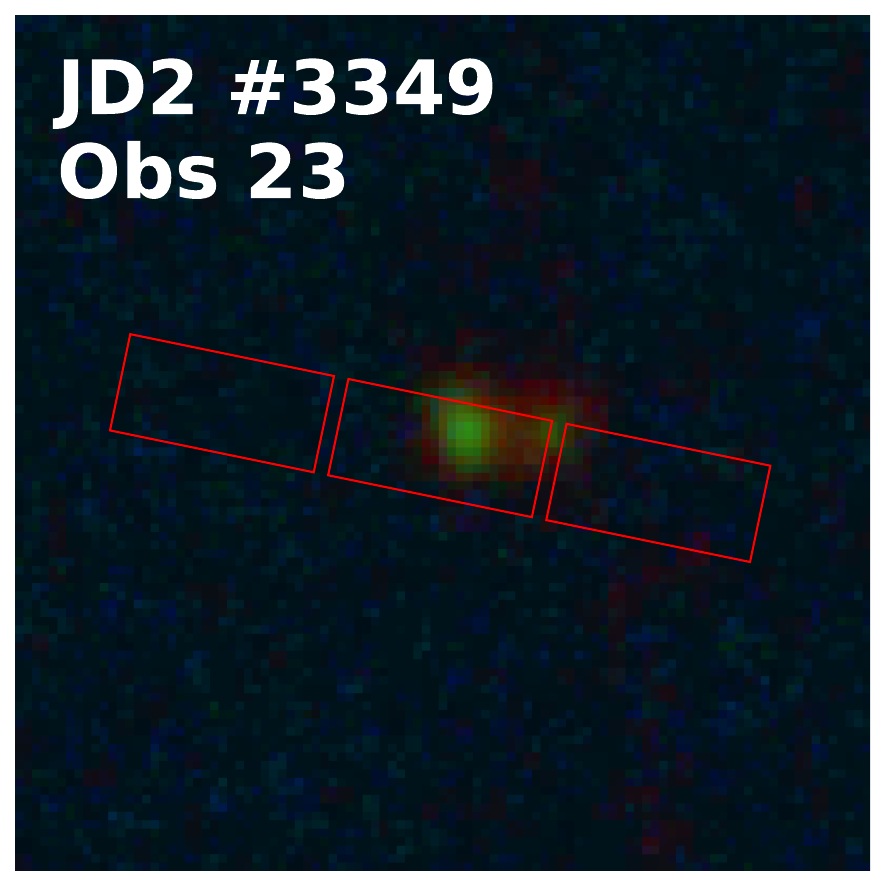}
\includegraphics[width=5in]{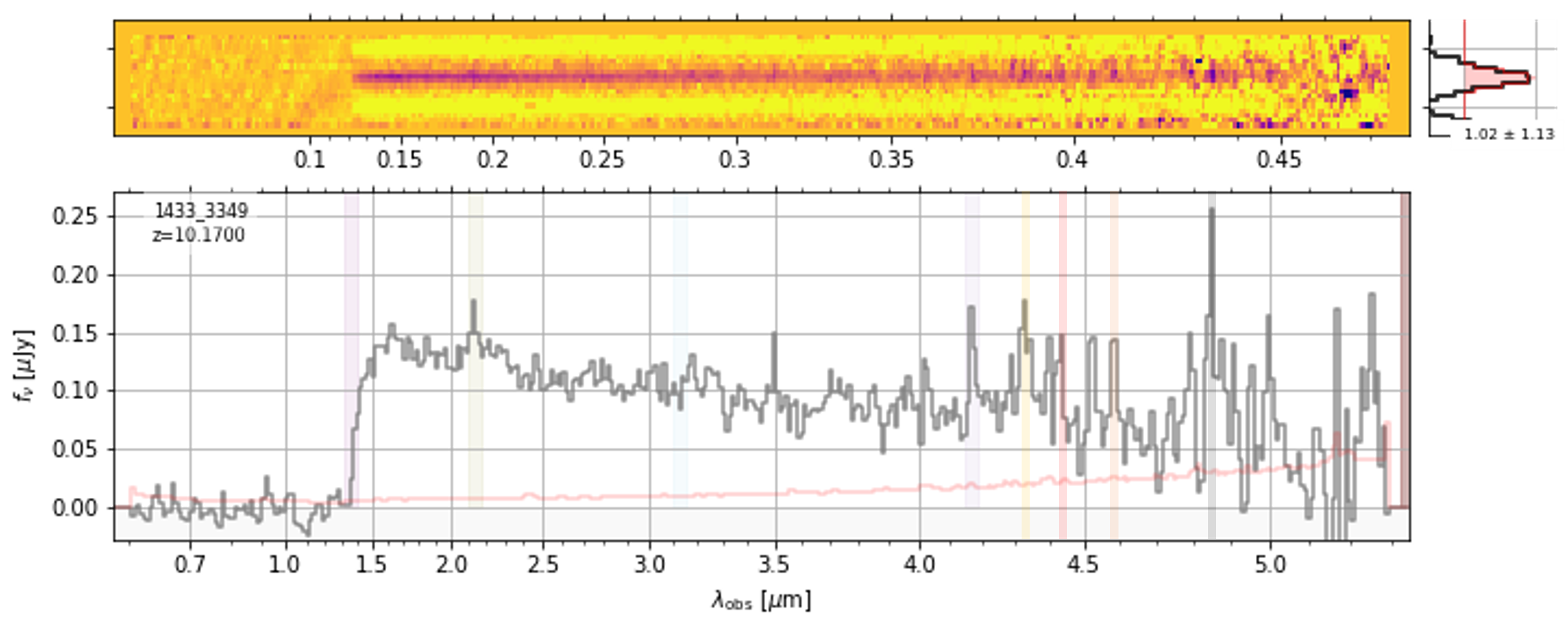}
\caption{Individual NIRSpec spectra of \JD\ lensed images JD1 and JD2 
(\#3593 and \#3349, respectively, in the catalog used to prepare NIRSpec Observations 21 and 23).
{\it Left}: Slitlets in red are overlaid on 2\arcsec $\times$2\arcsec\ NIRCam color images
(blue F115W; cyan F150W; green F200W; 
red F277W, F356W, F444W, F480M).
Note that most of the observations target the brighter component A.
{\it Right}: 2D spectra vs.~rest-frame wavelength for $z = 10.17$ (top)
and 1D spectra $F_\nu$ ($\mu$Jy) vs.~observed wavelength in microns (bottom).
In each 1D plot, the gray line shows the spectrum, and the red line shows the uncertainty.
Faint vertical colored shaded regions mark wavelengths of common emission lines:
\Lyaw\ (pink), \CIIIdw\ (green), \MgIIw\ (blue), \OIIw\ (purple){, \NeIIIw\ (yellow), \NeIIIwb\ (red), \Hdelta\ (orange), and \Hgamma\ (gray)}.
Only two of those (\CIIId\ and \OII) appear in \JD, along with others 
revealed in the {summed} spectrum (Figure \ref{fig:spectrum})
and listed in Table \ref{tab:lines}.
\label{fig:spectra}
}
\end{figure*}

\begin{figure*}
\includegraphics[width=\textwidth]{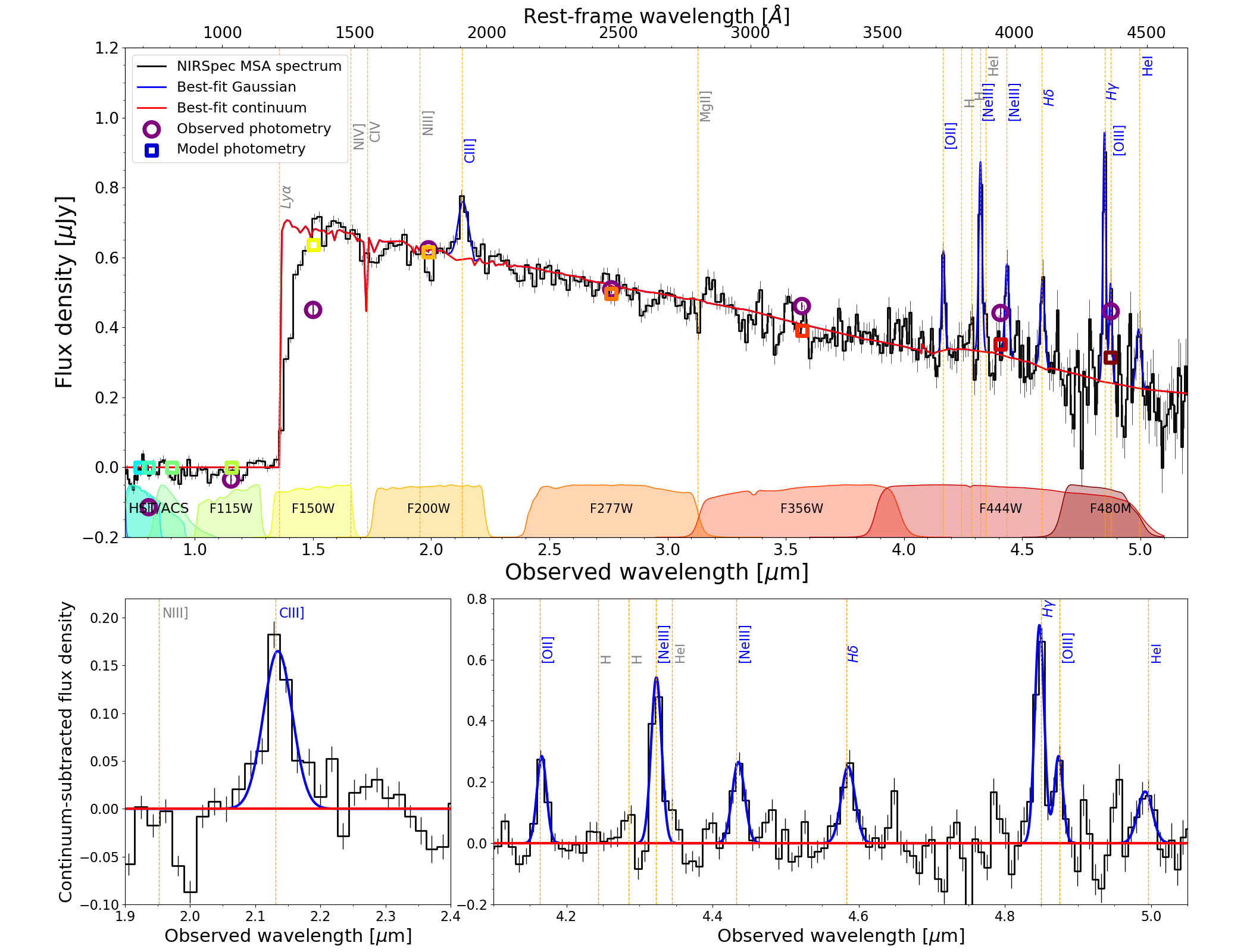}
\caption{Spectrum and photometry of \JD\ component A 
in the sum of both lensed images in both observations
(i.e., JD1 Obs 21 + JD1 Obs 23 + JD2 Obs 21 + JD2 Obs 23), with a total magnification \about $26.6 = 8+8+5.3+5.3$.
The spectrum is the {sum} of all 4 observed spectra.
The photometry shown here is measured within rectangular apertures 
resembling the NIRSpec slits in the 4 exposures. 
The red line is the \piXedfit\ model fit 
to the spectrum with emission lines masked out.
After subtracting this continuum (bottom panels), 
detected emission lines are fit with Gaussian functions (blue) to measure fluxes.
(Other undetected emission lines are labeled in gray.)
{The best-fit model continuum is smoothed with a constant spectral resolution (R) of 100, which corresponds to velocity resolution of $\sim$3000$\,$km/s.
We also see a hint of CIV absorption, as we mention in the last paragraph of \S\ref{sec:CIII}.
{Note that at the wavelength of CIV, $\sim1-2{\rm \mu m}$, R falls to $\sim30$, which is lower than R in our fitting.}}
\label{fig:spectrum}
}
\end{figure*}

\begin{figure*}
\centering
\includegraphics[width=\textwidth]{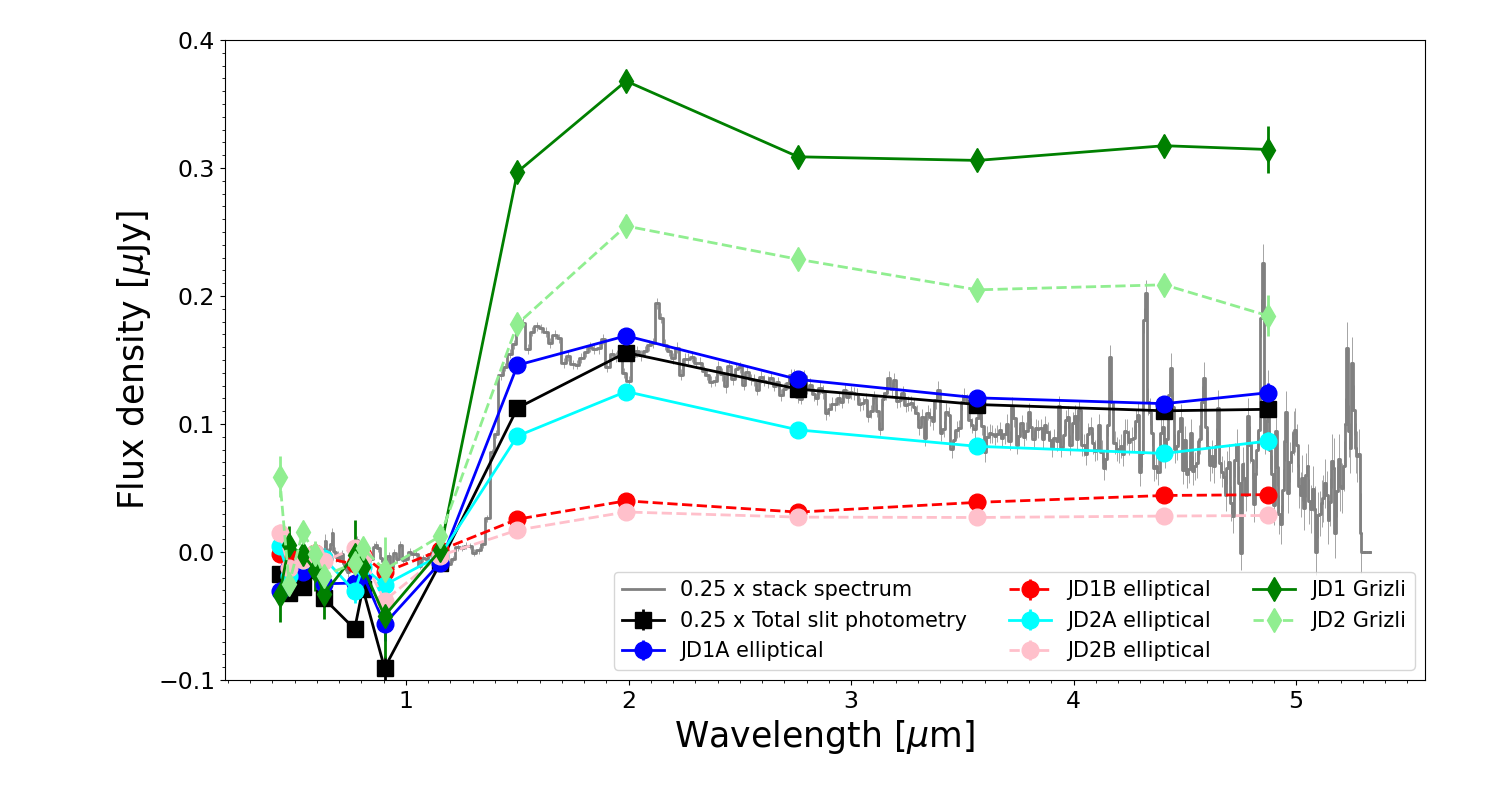}
\caption{\JD\ NIRSpec {summed} spectrum (divided by 4; gray)
compared with photometry presented in Table \ref{tab:pho}.
Black points show the photometry measured in 
rectangular apertures that mimic the observed NIRSpec slits.
These synthetic colors match the JD1A and JD2A
photometry measured within elliptical apertures by \piXedfit\ (blue and cyan).
The smaller clump B is redder (JD1B red, JD2B pink).
Photometry of the entire galaxy \JD\ lensed images JD1 and JD2 (dark and light green),
including clumps A and B plus additional diffuse light, was measured by \grizli\ with circular apertures of 0$\farcs$25 radius.
}
\label{fig:spectrum-photometry}
\end{figure*}

We measure \JWST\ photometry of the lensed images JD1 and JD2 using \grizli\ with circular apertures of 0$\farcs$25 radius.
{The segmentation process and photometry measurements with \grizli\ pipeline were done automatically to all detected galaxies in the field with circular apertures.}
We then measure the photometry of the JD components A and B individually 
using \piXedfit\ \citep{Abdurro'uf2023} within elliptical apertures as described in \citet{Hsiao2023} {centered at the central pixel (brightest pixel in F444W) of each component.}
{Both \grizli\ and \piXedfit\ described in \citet{Hsiao2023} v4 of the images while in this letter we anaylze v6 images as described earlier in this section.}
{No significant change is found in the result from SED fittings.}
Additionally, we create rectangular photometric apertures  that match the NIRSpec slitlets in size, position, and orientation in each of our 4 observations for JD1 and JD2 ({see} Figure \ref{fig:spectra})\footnote{\url{https://github.com/aabdurrouf/JWST-HST_resolvedSEDfits/tree/main/MACS0647-JD}}.
We find this photometry is very similar to the photometry of component A,
because the NIRSpec slitlets primarily targeted A
with occasional small contributions from the fainter clump B.
All resulting NIRCam photometry is presented in Table \ref{tab:pho} 
and plotted along with the NIRSpec spectrum in Figure \ref{fig:spectrum-photometry}.

We note the photometry of clump A {($\beta\sim-2.6\pm0.1$)} is slightly redder than the observed spectrum {($\beta\sim-2.8$)}.
Both measurements present challenges,
including the proximity of the A and B clumps,
and variable slit loss corrections as a function of wavelength
applied by the \JWST\ pipeline to the NIRSpec spectrum.


\input{observations}
\input{photometry}

\section{Results and Discussion}
\label{sec:results&discussion}

\input{line_flux}


We obtain four spectra of \JD\ component A 
in the lensed images JD1 and JD2 (Figure \ref{fig:spectra}).
Obs 21 of JD2 also includes flux from the fainter component B.
We have not attempted to extract signal from component B 
that might be isolated from A in the cross-dispersion direction.

The {summed} spectrum of all four observations 
(JD1 Obs 21, JD1 Obs 23, JD2 Obs 21, and JD2 Obs 23) 
is shown in Figure \ref{fig:spectrum}, with a total magnification of $26.6 = 8+8+5.3+5.3$.
%
We detect the continuum between 1.5 and 5.3\,\um\ with SNR $>$ 10 
per spectral resolution element out to \about 3\,\um;
flux decreases and noise increases to longer wavelengths.
Rather than a sharp Lyman break,
the spectrum rolls off gradually from 1.5$\,$\um\ (1343\AA\ rest-frame) 
to the Lyman break observed at 1.36$\,$\um\ (1216\AA\ rest-frame).
In \S\ref{sec:Lyadamping} we show that this attenuation is consistent with a {\Lya} damping wing.

We detect seven emission lines (including unresolved doublets): 
\CIIIdw, \OIIdw, \NeIIIw, \NeIIIwb, \Hdeltaw, \Hgammaw, and \OIIIwa, all with SNR $>$ 4. 
We also marginally detect \HeIwa\ with SNR $\sim$ 2.8.
Based on the observed wavelengths of these lines, 
we measure a spectroscopic redshift $z_{\rm spec}=10.17$. 
\cite{Harikane2023} recently reported the same redshift for \JD\ based on a fraction of the data: JD1 Obs 23.

The spectroscopic redshift is slightly lower (consistent within 2$\sigma$)
than the photometric redshifts estimated previously:
$z_{\rm phot}=10.7^{+0.6}_{-0.4}$ (95\% C.L.) from \HST\ imaging \citep{Coe2013} 
and $z_{\rm phot}=10.6\pm0.3$ from \JWST+\HST\ imaging \citep{Hsiao2023}.
This may be explained by the {\Lya} damping wing
reducing flux between 1.35 -- 1.5$\,$\um\ 
(in the \JWST\ F150W and \HST\ F140W filters)
more than expected for a sharp Lyman break, as assumed in the photometric redshift fitting
based on spectral energy distribution (SED) modeling.

We measure the emission line fluxes and report them in Table \ref{tab:lines}.
To measure the fluxes, we fit each line with a Gaussian 
after modeling and subtracting the continuum using \piXedfit\ \citep{Abdurro'uf2023}.
The continuum model is fit to the NIRSpec spectrum while masking out the emission lines.
The unresolved doublets \CIIId\ and \OII\ are each fitted with a single Gaussian.
Flux uncertainties are estimated using the Monte Carlo method.

Recently, \cite{Bunker2023} presented deeper NIRSpec spectroscopy\footnote{Exposure times of 
6.9 hours with the low-resolution ($R \sim 100$) prism and 
3.45 hours in each of three medium resolution ($R \sim 1000$) gratings: 
G140M/F070LP, G235M/F170LP, and G395M/F290LP} 
of GN-z11 at $z=10.603$
that featured all the lines above, 
plus additional lines that are {\it not} detected in MACS0647-JD:
\Lyaw, \NIVw, \CIVw, \HeIIw, \NIIIw, and \MgIIw.
The nitrogen lines \NIV\ and \NIII\ were surprising detections rarely observed in any galaxy
\citep{Cameron2023b,Charbonnel2023}.
The measured line fluxes were \about \tentotheminus{18} \cgsfluxunits.
Nitrogen lines that strong would have also been detected in our observations of \JD.
Note the continuum brightness of GN-z11 is comparable to JD1A (F200W AB mag {$25.8\pm0.1$}),
so similar line fluxes would correspond to similar equivalent widths
(generated by similar physical properties).


\subsection{Spectral modeling}
\label{sec:spectral}

We perform spectrophotometric SED fitting using \piXedfit\ \citep{Abdurrouf2021, Abdurrouf2022, Abdurro'uf2023}. This code fits the spectral continuum (with emission lines masked out) and photometry simultaneously. We use the {summed} spectrum and the total photometry measured with rectangular apertures resembling the NIRspec slits in the four exposures. This code uses Flexible Stellar Population Synthesis \citep[\texttt{FSPS};][]{Conroy2009, Leja2017} SED models, which have nebular emission incorporated using \textsc{Cloudy} \citep{Ferland2017}. We assume a \citet{Chabrier2003} initial mass function, \citet{2000Charlot} dust attenuation law, and the intergalactic absorption based on \citet{Inoue2014} model. 
The dust optical depth is denoted $\hat{\tau}_2$.
We assume a double power-law star formation history (SFH) with
rising slope $\beta$,
declining slope $\alpha$,
and timescale $\tau$:
SFR($t$) $= [(t/\tau)^\alpha + (t/\tau)^{-\beta}]^{-1}$.
Please refer to \citet[][Table 2 therein]{Abdurro'uf2023} for more information on the definition of these parameters and associated priors. 

Gas-phase metallicity \Zgas\ and stellar metallicity \Zstar\ are modeled independently.
To reduce the number of free parameters, 
we fix the ionization parameter $\log(U)=-2.0$ and the gas-phase metallicity \Zgas\ to \logOH\ $=7.5$ based on our estimate from emission lines (see \S \ref{sec:metallicity}, \S \ref{sec:ionization} and Table \ref{tab:phy}).
Varying these values within the uncertainties does not significantly change the results that follow.

First, we use \piXedfit\ to fit the spectral continuum after masking out the emission lines.
Then we subtract this best-fit continuum and fit Gaussians to the emission lines 
to measure their fluxes and equivalent widths, reported in Table~\ref{tab:lines}.

The spectrum has no Balmer break with $F_\nu (4200$\AA$) /F_\nu (3500$\AA$) \sim 0.77 \pm 0.06$,
measured as in \cite{Binggeli2019}, suggesting a young age.
Spectral continuum fitting suggests a very young mass-weighted age of $8^{+3}_{-2}\,$Myr for component A.
However, these spectral models do not account for the {\Lya} damping wing.
SED fitting to the photometry (\S\ref{sec:sed})
yields a similarly young age until we omit the F150W flux measurement affected by the Lyman damping.
Our measured photometry is also slightly redder than the spectrum, 
contributing to slightly older age measurements, though still young as we discuss next in \S\ref{sec:sed}.

\input{individual}

\subsection{Photometric SED modeling}
\label{sec:sed}

We perform spectral energy distribution (SED) fitting to the NIRCam photometry 
of the brightest lensed image JD1 (magnified by $\mu = 8 \pm 1$)
for the full galaxy and then for the individual clumps A and B,
as reported in Table \ref{tab:pho}.
We exclude the F150W photometry because our spectral models do not include {\Lya\ damping wing}.
(Including the F150W photometry results in artificially low ages.)
We perform SED modeling using 3 methods:
\piXedfit\ \citep{Abdurrouf2021, Abdurrouf2022, Abdurro'uf2023};
\bagpipes\footnote{\url{https://bagpipes.readthedocs.io}\label{fn1}} \citep{Carnall2018, Carnall2019b};
and \beagle\ \citep{Chevallard2016_BEAGLE}. 
We describe our methods in \S\ref{sec:spectral} for \piXedfit\ and below for \bagpipes\ and \beagle.
We summarize our results in Table \ref{tab:indi_phy}.

Our \bagpipes\ \citep{Carnall2018, Carnall2019b} method is described in detail in \cite{Hsiao2023}.
Briefly, we use Binary Population and Stellar Synthesis (BPASS) v2.2.1 templates \citep{Stanway2018} including binary evolution, 
with a \cite{Kroupa2001} IMF and upper mass cutoff of 300$\,$M$_\odot$. 
Nebular emission is included by processing these stellar models through the photoionization code \cloudy\ \citep{Ferland2017}. 
Dust attenuation is modeled with the \cite{Salim18} flexible parameterization,
including a variable slope and additional dust within star-forming regions $< 10$ Myr old.
Based on our spectroscopic results below, 
we fix gas metallicity to \Zgas\ = 0.1\Zsun\ and ionization parameter $\log(U)=-2$ {(see \S\ref{sec:ionization})}.
Varying these values within their uncertainties does not significantly affect our derived physical parameters.

We explore various star formation histories: constant, ``delayed $\tau$'', and non-parametric.
The delayed $\tau$ SFH allows for a linear rise followed by an exponential decline:
SFR($t$) $\propto t\, \exp(-t / \tau)$.
Most of our fits only have the linear rise with $\tau > 1$ Gyr.
We explore non-parametric star formation histories
with constant SFR in 4 time bins (e.g., 0 -- 10 -- 50 -- 150 -- 300 Myr).
With these models, we find almost all the SFR ends up in our youngest bin 0 -- 10 Myr,
with negligible older populations.

We also use \beagle\ \citep{Chevallard2016_BEAGLE}
to derive an independent fit to both the photometry and the measured emission line equivalent widths (see Table \ref{tab:lines}). 
\beagle\ assumes a \citet{Chabrier2003} IMF and also uses \cloudy\ to account for nebular emission. 
We further assume a delayed $\tau$ SFH with an ongoing starburst over the last 10\,Myr, an SMC dust attenuation law, and the \citet{Inoue2014} IGM attenuation models. 
We apply a similar setup and and parameter space as in \citet{Furtak2023}, 
though without the external priors and fixing the redshift to the measured spectroscopic value.
Here we also leverage the additional information provided by the numerous emission lines 
to also leave the gas-phase metallicity and the dust-to-metal ratio as free parameters. 
These are modeled with uniform priors $\log(Z_{\mathrm{gas}}/\mathrm{Z}_{\odot})\in[-2.2, -0.3]$ and $\xi_{\mathrm{d}}\in[0.1,0.5]$ {as in \citet{Gutkin2016} and \citet{Chevallard2016_BEAGLE}}.

Based on these various methods, we obtain results summarized in Table \ref{tab:indi_phy}.
We find \JD\ has a stellar mass \logM\ $= 8.1 \pm 0.3$;
SFR $10 \pm 3$ \Msun\ yr\inv;
and high sSFR  \about\ 80 Gyr\inv,
where the specific star formation rate sSFR = SFR / stellar mass.

The clump A photometry can be well fit by 
constant star formation for $17^{+5}_{-4}$ Myr,
corresponding to mass-weighted ages $8^{+3}_{-2}$ Myr,
according to \bagpipes.
Age estimates from the various methods range between 1 -- 50 Myr, 
consistent with the results from \cite{Hsiao2023}.
We estimate clump A contains about 1/3 the stellar mass of \JD.

The smaller redder component B photometry is best fit by a linearly rising SFR for 40 -- 300 Myr,
with a mass-weighted age 10 -- 110 Myr,
stellar mass \logM\ $= 7.9\pm0.3$,
and dust $A_V = 0.10 \pm 0.05$ mag.
Based on similar age estimates in \cite{Hsiao2023},
they suggested clump B may have formed separately and merged with the younger clump A,
though different star formation histories in situ are also possible.


\JD\ has an absolute UV magnitude $M_{UV} = -20.3 \pm 0.2$,
including $M_{UV} = -19.5 \pm 0.2$ for clump A
and $M_{UV} = -17.9 \pm 0.2$ for clump B.
These are based on the JD1 F200W photometry in Table \ref{tab:pho} delensed by a magnification $\mu = 8 \pm 1$.

\input{line_ratios}

\subsection{Star formation rate and history}

Hydrogen Balmer line strengths yield robust estimates of recent star formation history 
within the past $\lesssim$ 10 Myr.
These measurements commonly utilize 
the strongest Balmer line \Halpha\ or the second strongest \Hbeta\ {\citep[e.g.,][]{Glazebrook1999,Erb2003,Reddy2008,Madau2014,Shapley2023}}.
Both of these are redshifted beyond our observed wavelength range,
however we detect both \Hgamma\ and \Hdelta.

Intrinsic Balmer line strength ratios are dictated by atomic physics.
For Case B recombination at $T=10000\,$K, we expect \Hgamma\ / \Hdelta\ $= 1.81$ \citepeg{Dopita2003,Groves2012}.
In MACS0647$-$JD, we measure \Hgamma\ / \Hdelta\ $\sim1.88\pm0.43$, 
consistent with expectations (\Hgamma\ / \Hdelta\ $= 1.81$) within the uncertainties.
Dust reddening may artificially increase observed Balmer line ratios.
We cannot place meaningful constraints on the dust extinction,
given the large uncertainties in the flux ratio 
and the similar wavelengths of \Hgamma\ and \Hdelta.
Note that Balmer line stellar absorption might affect our measurement of \Hgamma\ and \Hdelta\ of $0-5\,$\AA, which is within our uncertainty of $6\,$\AA. 
We assume no dust, consistent with the result $A_V < 0.02$ mag 
from SED fitting to the photometry of component A in \cite{Hsiao2023}
and similarly in our updated analysis of the latest photometry \S\ref{sec:sed}.
Finally, we note the measured \Hgamma\ rest-frame equivalent width EW 42$\pm$6 \AA\ may be
slightly suppressed by photospheric absorption lines with EW $<5$\AA,
but that is within our measurement uncertainty.

Assuming no dust, we estimate the \Hbeta\ and \Halpha\ line fluxes using the 
expected line ratios \Hbeta\ = 2.14 \Hgamma\ and \Halpha\ = 6.11 \Hgamma\ (see Table \ref{tab:ratio}).
Converting our inferred \Halpha\ line flux to luminosity, 
including a de-lensing correction for magnification,
we estimate 
$L_{\rm H\alpha}=(4.5\pm0.2)\times10^{41}\,{\rm erg\,s^{-1}}$. 
Adopting an SFR conversion factor of 3.2 $\times$ \tentotheminus{42} 
suitable for high-redshift galaxies \citep{Reddy2018},
we estimate SFR $=1.4\pm0.2\,$\Msun\ yr\inv.
Note that the SFR will be higher if there is dust,
and the conversion factor is sensitive to the IMF, star formation history, and metallicity assuming that the escape fraction is zero \citepeg{Wilkins2019}.
{Also, the SFR derived from the emission lines is smaller than the estimates from the photometry of JD1, but it is consistent with the SFR estimated from the photometry of JD1A.
This is likely because the slitlets of four observations mainly targeted JDA (see Figure \ref{fig:spectra}).}

Similarly, \OIIw\ can also be used to estimate the SFR for distant galaxies \citep[e.g.,][]{Kennicutt1998}.
This yields a slightly lower ${\rm SFR_{OII}}=0.6\pm0.2\,$\Msun\ yr\inv.
However, we note that this empirical calibration was derived for low-redshift galaxies and depends on metallicity and dust,
which are lower for high-redshift galaxies like \JD.


\subsection{Strong \CIIIdw\ emission}
\label{sec:CIII}

We detect strong emission of \CIIIdw\ (blended) with a line flux of 
$(3.1\pm0.6)\times10^{-18}$ erg s$^{-1}$ cm$^{-2}$
and rest-frame equivalent width 14$\pm3\,$\AA.
Such high equivalent width \CIIIdw\  is associated with low metallicity galaxies experiencing an intense burst of star formation \citep{Rigby2015}; 
it is predicted for BPASS models {\citep{Stanway2018}}
only when they include very recent star formation within the past Myr
\citep{Jaskot2016}, 
consistent with our young age estimates \S\ref{sec:sed}.
We use the measured equivalent width as a metallicity diagnostic in \S\ref{sec:metallicity}.

We note additional \CIIId\ emission appears to come from 
the B component, which is marginally spatially resolved 
in the 2D spectrum of JD2 in Obs 21 (Figure \ref{fig:spectra}).
This would suggest a young age for B,
or at least very recent star formation {within the past Myr}
perhaps in addition to the older star formation and age (\about 100 Myr)
estimated from the photometry.

Strong \CIIId\ emission with high rest-frame equivalent widths of 10 -- 20$\,$\AA\ or more 
has been observed in $z \sim 0$ -- {8} galaxies,
often associated with high observed or inferred
\OIII+\Hbeta\ equivalent widths \about\ 1000\AA\ -- 2000\AA\ \citepeg{Stark2017,Berg2019,Hutchison2019,Mainali2020,Tang2021,Mingozzi2022, vanzella_ion2_2020, vanzella_sunrise_2023} 
or even higher than $2000\,$\AA\ \citep{Tang2023}.
This is consistent with our expectation
based on our extrapolated \OIII+\Hbeta\ flux (Table \ref{tab:ratio}),
which roughly corresponds to an equivalent width \about\ 1000\AA,
assuming the continuum flux at rest-frame \about 5000\AA\ is similar to 
what we measure out to \about 4500\AA.

We do not detect the higher ionization species \CIVw.
This combined with the strong \CIIId\ emission suggests that 
an AGN is likely not the driving ionizing source in this galaxy.
Detections of strong \CIV\ are often associated with AGN activity \citepeg{Baskin2005,LeFevre2019}.
However, young stellar populations can also drive stronger \CIV\ emission as well \citepeg{Mainali2017}.

If anything, we see a hint of \CIV\ absorption,
as also predicted by our \piXedfit\ best-fit model.
Such \CIV\ absorption is observed (and modeled) for young stellar populations with ages $<$ 10 Myr
\citepeg{Leitherer1999,Chisholm2019}
and can constrain stellar age and metallicity when detected at high signal-to-noise and resolution.
The ISM can also absorb \CIV\ due to internal gas kinematics and/or higher column density \citep{Steidel2016,Jaskot2016}.
\subsection{Ionization \logU\ and \logxiion} 
\label{sec:ionization}

The line ratio O32 $=$ \OIIIw\ $/$ \OIIw\ has long been used 
\citepeg{Hicks2002}
as a diagnostic for the ionization parameter \logU.
At high redshifts, when \OIIIw\ is unavailable {(For example, galaxies at $z\gtrsim9.6$, \OIIIw\ shifts beyond the wavelength coverage of 5.3$\,{\rm \mu m}$ in NIRSpec)},
the line ratio Ne3O2 $=$ \NeIIIw\ $/$ \OIIw\ has been
proposed as an alternative diagnostic
\citep{Perez-Montero2007, Levesque2014, Maiolino2019, Witstok2021}.
Strong correlations are observed between Ne3O2 and O32
in nearby galaxies when all lines are detected with strong significance
\citep{Jaskot2013,Izotov2018,Flury2022}, especially the case for higher ionization (i.e., similar to properties of galaxies expected and found at $z>6$).
More distant galaxies at $z \sim 2$ -- 7 exhibit similar trends
\citep{Christensen2012a,Tang2019,Vanzella2020,Tang2023}.

In \JD, we measure a high value for Ne3O2 $= 1.8 \pm 0.2$.
According to \citet[][Equations 1--3]{Witstok2021},
this Ne3O2 ratio corresponds to O32 $ = 30 \pm 6$
and ionization parameter \logU\ $= -1.84 \pm 0.06$,
using \logU\ = 0.80 $\times$ log(O32) $-$ 3.02 from \citep{Diaz2000}.


We also used \HIIC\ version 5.22\footnote{
\href{https://home.iaa.csic.es/~epm/HII-CHI-mistry-opt.html}
{https://home.iaa.csic.es/\about epm/HII-CHI-mistry-opt.html}} \citep{Perez2014} to estimate \logU\ using the observed and extrapolated line flux ratios discussed in \S\ref{sec:metallicity} and listed in Table \ref{tab:metal}. 
Our inputs were measured line fluxes for \OIIw, \NeIIIw, and \OIIIwa, all normalized to our extrapolated \Hbeta\ flux based on \Hgamma.
Here we did not use the extrapolated \OIIIw\ flux.
We obtain an ionization parameter of \logU\ $= -2.09\pm0.02$,
similar to but slightly lower than the value estimated above from Ne3O2.

\cite{Kewley2019} noted the line ratio \CIIIdw\ $/$ \CIIw\ could be
an excellent diagnostic for ionization parameter \logU\ at low metallicities \logOH\ $< 8.5$
(see their Figure 7).
Our non-detection of \CII\ with flux $< 0.55$ \en{18} \cgsfluxunits\ (1\,$\sigma$ limit),
yielding a line ratio \CIIId\ $/$ \CII\ $> 5.7$ 
corresponds to \logU\ $> -2.8$ for \logOH\ \about\ 7.6.

We conclude \logU\ $= -1.9\pm0.2$ based on the three estimates (see also Table \ref{tab:phy}).
This is comparable to the value \logU\ \about\ $-2$ obtained for GN-z11 \citep{Bunker2023}.

Our \Halpha\ luminosity measurement (delensed and assuming no dust)
yields an estimate for the production rate of hydrogen-ionizing photons:
\Ndotion\ = (3.3$\pm0.1$)\e{53} s\inv (${\rm log}(L_{\rm H\alpha})={\rm log} (N_{\rm ion}/{\rm s})-11.87$).  
{Assuming the escape fraction is zero}, combined with our measured and demagnified UV luminosity for clump A
(corresponding to $M_{UV} = -19.5 \pm 0.2$),
we estimate an ionizing photon production efficiency 
$\xi_{\rm ion} = \dot{N}_{\rm ion} / L_\nu^{\rm UV} = 1.6\pm0.3$ \e{25} erg\inv\ Hz,  
or log($\xi_{\rm ion})=25.2\pm0.2$.  
This is consistent with predictions for high-redshift galaxies ($z>8$), 
with higher values expected for younger {($\lesssim10^{8}\,{\rm yr}$)}, lower-metallicity galaxies {($Z\lesssim Z_{\odot}$)} 
\citep[e.g.,][]{Schaerer2003,Robertson2013,Wilkins2016}.
Observationally, recent \JWST\ studies reported higher values 
${\rm log}(\xi_{\rm ion})\sim25.7-26.0\,$erg\inv\ Hz 
for high-redshift ($7<z<11$) galaxies
\citepeg{Tang2023,Bunker2023,Fujimoto2023}.
Our \Ndotion\ estimate should be treated as a lower limit
that could increase due to dust and/or escape fraction \fesc.


\input{metallicites}

\input{properties}

\subsection{Metallicity}
\label{sec:metallicity}

We use various methods and line ratio diagnostics
to estimate \JD's gas-phase metallicity
\logOH\ = 7.5 -- 8.0, or (0.06 -- 0.2) \Zsun,
where we adopt a solar metallicity \logOH\ = 8.69 \citep{Asplund2021}.
Table \ref{tab:metal} summarizes our results, and we describe our methods below.

The auroral line \OIIIwa\ can provide a ``direct'' metallicity measurement 
via electron temperature when detected and measured along with other lines 
including \OIIIww\ \citepeg{Peimbert1967,Osterbrock1989,Sanders2023}.
Given our estimate O32 $ = 30 \pm 6$ (\S\ref{sec:ionization})
and observed line flux for \OII, 
we can estimate the expected \OIIIw\ line flux as 
$(1.3\pm0.3)\times10^{-17}\,$\cgsfluxunits, assuming no dust,
or \about 50 times stronger than \OIIIwa.
Note the \OIIIw\ line is 2.98 times stronger than \OIIIwc\ \citep{StoreyZeippen2000},
so our extrapolated combined \OIIIww\ flux is 1.34 times the \OIIIw\ flux.


We use \pyneb\ \citep{pyneb_Luridiana2015} to analyze the observed and extrapolated emission line ratios.
For the entire procedure, we assumed no dust and an electron density $n_{\rm e}=100$~cm$^{-3}$.
First, we use the task \texttt{getCrossTemDen}, 
taking into account the \OIIIw\ $/$ 4363 \about\ 50 with a 20\%\ uncertainty.
In this way, we obtain the electron temperature of the high-ionization zone $T_{\rm e}$(\OIII) $= 15000 \pm 1400$~K. The $T_{\rm e}$ uncertainty is estimated by running the task 1000 times and taking the standard deviation of all the values. 
Then, we use the \citet{Garnett1989} relation to estimate the low-ionization 
 gas temperature $T_{\rm e}$(\NII) $= 13600 \pm 1000$ K.
Finally, we calculated the O$^{++}$ and O$^{+}$ ionic abundances with the task \texttt{getCrossTemDen}, using as input \OIIIw\ $/$ \Hbeta\ \about\ 8 and $T_{\rm e}$(\OIII), and \OIIw\ $/$ \Hbeta\ \about\ 0.26 and $T_{\rm e}$(\NII), respectively, with the same method to estimate the uncertainties.
The total of both species yields \logOH\ $= 8.0 \pm 0.2$,
or \about 20\% \Zsun\ (0.13 -- 0.32 \Zsun).

We also utilize 
\HIIC\ version 5.22 to analyze our observed line ratios, as described in \S\ref{sec:ionization}. 
We obtain \logOH\ \about\ $7.5\pm0.2$, corresponding to \about $0.06^{+0.04}_{-0.02}$\Zsun.



Next, we turn to line ratio diagnostics, beginning with Ne3O2.
Previous reviews find no strong correlation between Ne3O2 and metallicity,
as seen, for example, in local metal-poor galaxies {($Z<0.6\,Z_{\odot}$)} analyzed by
\citet[][Figure 9 therein]{Nakajima2022}.
But the few galaxies with high Ne3O2 \about\ 2 in that study
were clustered between roughly 
\logOH\ $= 7.6 \pm 0.2$,
or 0.05 -- 0.13 \Zsun.
It is unclear if that low-redshift result {($z<0.1$)} can be extended to high-redshift galaxies.

Another metallicity diagnostic is the line ratio R23 = (\OIIIww + \OIIw) $/$ \Hbeta\ \citep[e.g.,][]{Nakajima2022,Sanders2023}.
Based on our observed \OII, extrapolated \OIII, and \Hbeta\ inferred from \Hgamma, 
we extrapolate R23 $= 11 \pm 3$.
This is among the highest values measured for local metal-poor galaxies by \cite{Nakajima2022}.
That puts R23 \about\ 11 in
the zone of confusion in the bimodal R23--metallicity relation,
severely limiting its usefulness as a metallicity diagnostic,
though it can still provide a rough estimate.
We note in \citet[][Figure 3]{Nakajima2022},
that galaxies with R23 $ = 11 \pm 3$ have metallicities
\logOH\ $= 8.0 \pm 0.3$,
or between \about 0.1 and 0.4 \Zsun.
We note some of the scatter in the R23 vs.~$Z$ is due to ionization,
with higher ionization yielding higher R23 values for a fixed metallicity
\citep[e.g.,][Figure 5]{Perez2014}.


\cite{Izotov2021} propose a metallicity indicator combining R23 and O32 (their Equation 5)
for improved accuracy at low metallicities $<$ 6\% \Zsun, or \logOH\ $< 7.5$.
\cite{Nakajima2023} show this relation works up to \logOH\ \about\ 7.8
for $z = 4$ -- 8.5 galaxies with direct metallicity measurements from NIRSpec.
For MACS0647-JD, this relation yields 12 + log(O/H) \about\ $7.7\pm0.2$, 
or 0.06 -- 0.16 \Zsun.


Finally, we use the rest-frame UV line \CIIIdw\ as a metallicity diagnostic.
\cite{Mingozzi2022} analyzed 45 local ($z<0.2$) high-redshift analogs observed by the CLASSY survey
\citep{Berg2022_CLASSY},
measuring a relation between \CIIId\ EW and metallicity with 0.18 dex intrinsic scatter
for galaxies with EW $>$ 5\AA.
Using their equation 7 and our observed \CIIId\ EW 14.2$\pm2.5\,$\AA\ (discussed in \S\ref{sec:CIII}),
we estimate \logOH\ $= 7.84 \pm 0.18$, or 0.09 -- 0.21 \Zsun.

We find broad general agreement among our 6 metallicity indicators
with best-fit estimates ranging between 
\logOH\ = 7.5 -- 8.0, or (0.06 -- 0.2) \Zsun,
as summarized in Table \ref{tab:metal}, which is also consistent with the metallicity of \logOH\ = 7--7.8 expected from Astraeus simulations for galaxies at z$\sim$10 with a stellar mass of $10^{8.1}\,M_{\odot}$ \citep{Ucci2023}.
This is promising for future studies that only have one or more of these diagnostics available
either in the rest-frame optical or rest-frame UV.
On the other hand, greater precision will be achieved by generally improving these diagnostics
and in our specific case by obtaining additional spectra of \JD\ covering longer wavelengths
including \OIIIww, \Hbeta, and \Halpha.



\begin{figure*}
\includegraphics[width=\textwidth]{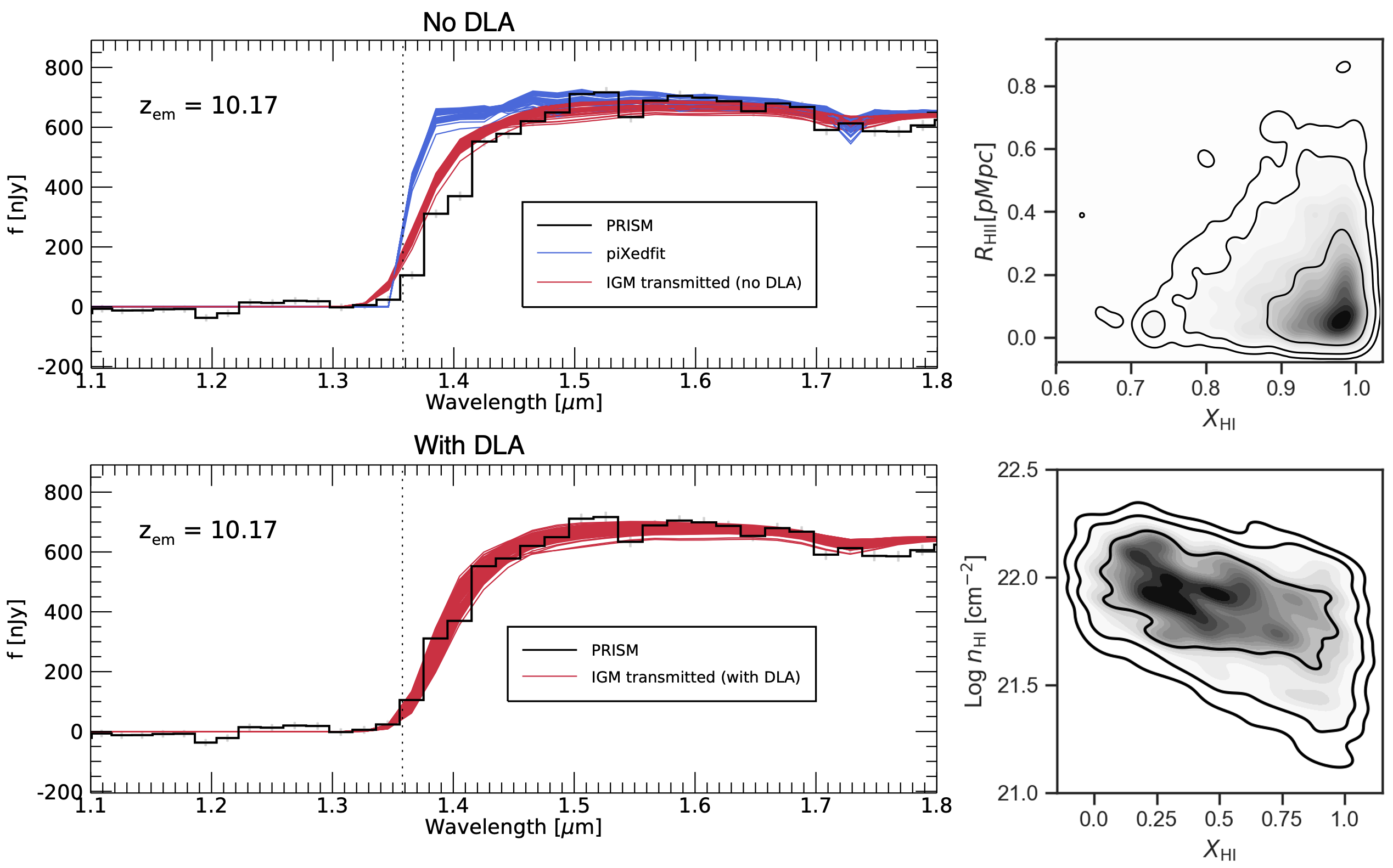}
\caption{\label{fig:Lyman-damping}
{{\it Top:} \Lya\ damping wing analysis of MACS0647-JD without a damped \Lya\ (DLA) contribution.  The left panel shows the \piXedfit\ SED model (blue) convolved with the IGM transmission and NIRSpec prism spectral resolution, resulting in the transmitted model spectrum (red). The vertical dotted line denotes the systemic wavelength of \Lya\ at $z = 10.17$. The right panel displays 1$\sigma$, 2$\sigma$, and 3$\sigma$ contours of the probability distribution of $X_{\rm HI}$ and $R_{\rm HII}$. The best fit suggests a highly neutral IGM ($X_{\rm HI} > 0.9$ at a 1$\sigma$ lower limit) with a small ($R_{\rm HII} < 0.2$ pMpc) ionized bubble around \JD\, although the modeled spectra show a still considerably steeper break than observed (black). {\it Bottom:} \Lya damping wing analysis with a DLA system. The inclusion of a DLA contribution better reproduces the observed \Lya break shape (left panel). However, the \ion{H}{1} column density of the DLA ($n_{\rm HI}$) is highly degenerate with the neutral IGM fraction, making it highly unconstrained.}}
\end{figure*}

\subsection{\Lya\ damping wing}
\label{sec:Lyadamping}

The {\Lya} photon scattering cross section as a function of wavelength is given by the Voigt profile, which combines a Gaussian core and Lorentzian wings \citep{Tasitsiomi2006}. Resonant scattering of {\Lya} photons by the neutral IGM suppresses line and continuum fluxes at wavelengths bluer than the rest-frame {\Lya} at 1216\AA.
Additionally, the Lorentzian wing cross section accumulates the attenuation to the redward continuum, softening the shape of {\Lya} break.
This is known as {\Lya} damping wing attenuation {\citep{MiraldaEscude1998}}.
{Features like these have been observed in quasars \citep[e.g.,][]{Mortlock2011, Banados2018, Davies2018, Wang2020, Yang2020}, and they have been utilized to investigate not only the average neutral fraction but also specific regions along different sightlines in the IGM at $z>7$.}
Recently, this effect has been observed in NIRSpec prism observations of high-redshift galaxies {
\citep[$z>8$;][]{Bunker2023, Boyett2023, Arrabal-Haro2023a, Arrabal-Haro2023b, CurtisLake2023} }

We model the {\Lya} damping wing optical depth following Equation (30) in 
\cite{Dijkstra2014}, as done in \cite{Arrabal-Haro2023a, Arrabal-Haro2023b}.
The {\Lya} damping wing description has two free parameters 
(with redshift held fixed at our measured $z = 10.17$): 
(1) the neutral hydrogen fraction in the IGM ($X_{\text{HI}}$) 
and (2) the size of the ionized bubble ($R_{\text{HII}}$), or Stromgren sphere.
In short, a higher neutral fraction in the IGM makes the break cutoff become smoother, 
and a larger ionized bubble size shifts the break blueward 
allowing the escape of photons at shorter wavelengths than the rest frame {\Lya}.
To reproduce the observed continuum spectrum, 
the galaxy model SEDs are convolved with the IGM transmission 
($T_{\text{IGM}}\propto e^{-\tau}$, where $\tau$ is damping wing optical depth).
We then smooth the model spectra with prism spectral resolution 
and apply an additional 5\% of systemic errors in the $\chi^{2}$ estimation.

We perform fitting using an MCMC sampler package described in \cite{Jung2017}.
We use flat linear priors for the free parameters \XHI\ and \RHII.
In each MCMC chain step, we take a random model spectrum from the range of models generated by \piXedfit\ (see \S\ref{sec:sed}) 
to accommodate the uncertainties in galaxy SED modeling.
{The wavelength range of the spectrum fitted by \piXedfit\ is between $1.5-5.2\,{\rm \mu m}$ and the fitting does not include the modeling of the absorption by Damped Lyman-$\alpha$.
The inclusion of the additional attenuation by an immediate Damped Lyman-$\alpha$ can reproduce the observed spectra well as in \citet{Heintz2023}.}

Figure \ref{fig:Lyman-damping} shows
our results suggesting a small size of an ionized bubble 
(\RHII\ $\ll 1$ pMpc)
with a highly neutral IGM (\XHI\ $>$ 0.9 at a 1$\sigma$ lower limit).
{We experiment with different priors on \XHI, 
including simply assuming a completely neutral IGM (\XHI\ $=1$)
as inferred from other work \citep[e.g.,][]{Barkana2001,Treu2013,Mason2018,Bruton2023}, and a logarithmic prior.
These priors do not significantly affect our bubble size estimates.
}

The $R_{\text{HII}}$ {estimate} appears reliable in the sense that the observed spectra show no positive signals blueward of the systemic wavelength of {\Lya}.
However, the interpretation of the $X_{\text{HI}}$ measurement is rather complicated. 
The modeled spectra present a {\Lya} break still considerably steeper than the observed spectrum.  
As the modeled spectra already include {\Lya} damping wing opacity with almost fully-neutral IGM, such discrepancies cannot be explained with the IGM attenuation alone.
In fact, the total {\Lya} optical depth combines the IGM damping wing optical depth ($\tau_{\rm IGM}$) and the optical depth due to resonant scattering within the  circumgalactic \citep[CGM;][]{Dijkstra2014}. 
However, the latter part is missed in our simple description of damping wing opacity. 
To mitigate the differences between the modeled spectra and the observed, it is necessary to carefully model the CGM contribution \citep{Sadoun2017, Weinberger2018, Mason2020}.  
Including the CGM contribution would result in additional softening on {\Lya} break closer to the observed spectrum, which possibly requires less contribution from IGM attenuation and could allow for a slightly more ionized IGM. 

{We attempt to model the spectra with a damped Lyman-alpha (DLA) system, following the approach in \citet{Heintz2023}. The results with the DLA contribution are shown at the bottom of Figure \ref{fig:Lyman-damping}. The inclusion of a DLA better reproduces the observed spectrum, with a DLA \ion{H}{1} column density of ${\rm log}(n_{\rm HI},{\rm cm^{-2}})=21.9\pm0.2$, comparable to \citet{Heintz2023}. This analysis reveals that the IGM damping wing attenuation is highly degenerate with the effect of DLA systems. Consequently, including DLAs makes the neutral hydrogen fraction highly unconstrained (bottom right in Figure \ref{fig:Lyman-damping}). However, this improved fit underscores the complexity of disentangling the contributions from DLAs and the IGM in such high-redshift observations. However,} detailed modeling of {\Lya} attenuation of the CGM requires knowledge of density profiles of H{\sc I} gas and radiation field from a host galaxy \citep{Weinberger2018, Mason2020}, which is beyond the scope of this study. 
Also, the relation between IGM attenuation and bubble size can deviate from the theoretical model of \cite{Dijkstra2014} due to the stochasticity of individual sightlines, such as fluctuations in density/velocity and random encounters of self-shielded systems \citep{Hutter2014, Park2021, Smith2022}. 
Ideally, one needs to average over a large number of observations to suppress such uncertainties. 
Additionally, the uncertainties in modeling intrinsic rest-UV spectra make the interpretation more complicated.

\subsection{Ionized bubble size}
\label{sec:bubblesize}

We also estimate the radius of the ionized bubble around \JD\ by
integrating the total ionizing photon production over the galaxy's lifetime
as in \cite{Shapiro1987}:

\begin{equation}
\frac{4}{3} \pi R_{\rm HII}^3 \sim 
\frac{\dot N_{\rm ion}\, f_{\rm esc}\, t_{*}}{X_{\rm HI}\, n_{\rm H}}.
\end{equation}


We estimate \Ndotion\ \about\ 3.3\e{53} s\inv\ based on SFR \about\ 1.4 \Msun\ yr\inv\ for clump A
(\S\ref{sec:ionization})  
with stars having formed continuously for perhaps $t_*$ \about\ 20 Myr (\S\ref{sec:sed}).
The average density of hydrogen particles in the universe at $z = 10.17$ 
is $n_{\rm H}$ \about\ 7\e{69} Mpc$^{-3}$, where 
$n_{\rm H}=(1-Y_{\rm He})\,\rho_{\rm crit}\,\Omega_{\rm b}\,(1+z)^3\, m_{\rm H}^{-1}$.
{And here we assume a neutral IGM ($X_{\rm HI}=1$).}

We do not estimate $f_{\rm esc}$ for \JD\ in this work.
Here, it suffices to explore possible values.
For example, $f_{\rm esc} = 10$\% yields \RHII\ \about\ 0.09$\,$pMpc.
A more extreme $f_{\rm esc} = 50$\% would only increase that to \RHII\ \about\ 0.15$\,$pMpc.


Above, we use the SFR for clump A. 
The entire galaxy has SFR \about\ 10 \Msun\ yr\inv, or \about\ 7$\times$ higher.
We might expect that most of the ionizing radiation comes from the blue clump A.
But even generously assuming the same 
ionizing photon production efficiency \xiion\ for the entire galaxy
and increasing \Ndotion\ by 4$\times$,
\RHII\ only increases by a factor of \about 1.6
to \RHII\ \about\ 0.24 pMpc, again assuming a very high $f_{\rm esc} = 50$\%.

{
We note these are only rough estimates subject to significant uncertainties.
Dense neutral gas clouds in the IGM may impede the expansion of the reionized bubble.
Some of the reionized gas may recombine, further reducing the bubble size,
especially if the ionizing flux wavers due to variations in SFR or \fesc.
}

{On the other hand, the total ionizing flux might be much stronger than we have estimated here 
due to fainter undetected companions, especially given that \JD\ is a merger.}

\begin{figure*}
\includegraphics[width=2in]{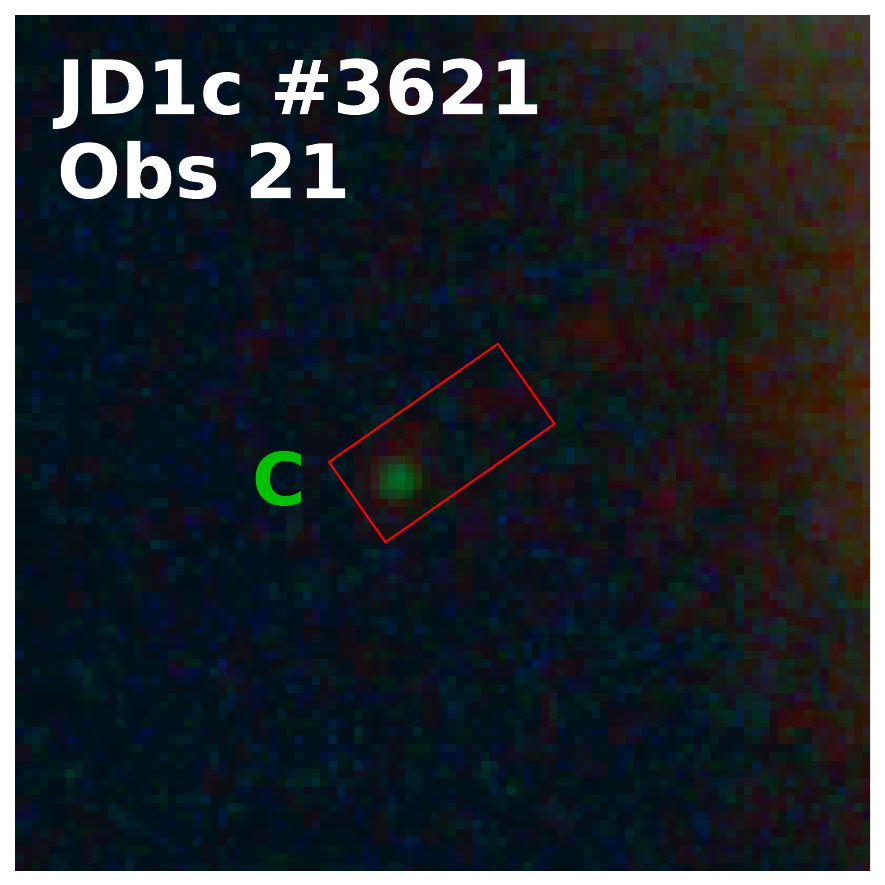}
\includegraphics[width=5in]{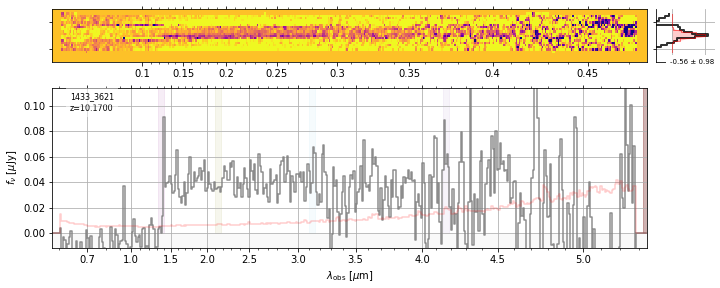}
\includegraphics[width=2in]{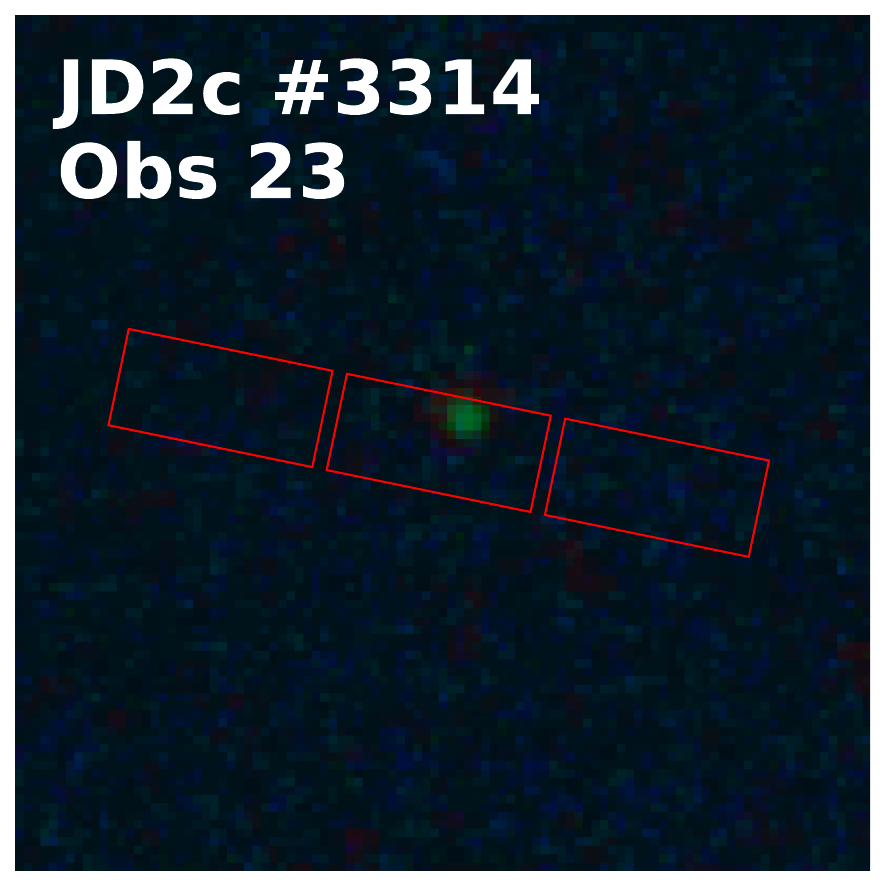}
\includegraphics[width=5in]{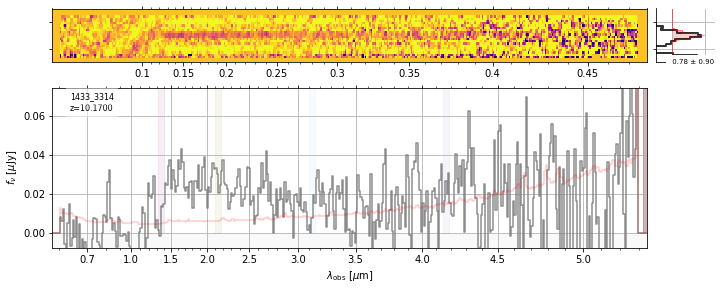}
\caption{NIRSpec prism spectra of two lensed images of the 
candidate companion galaxy C.
Figures show slitlet coverages, 2D spectra, and 1D extractions, 
similar to Figure \ref{fig:spectra} for \JD.
Redshift is fixed to the same value $z = 10.17$.
\label{fig:JDC_spectra}
}
\end{figure*}

\begin{figure*}
\includegraphics[width=\textwidth]{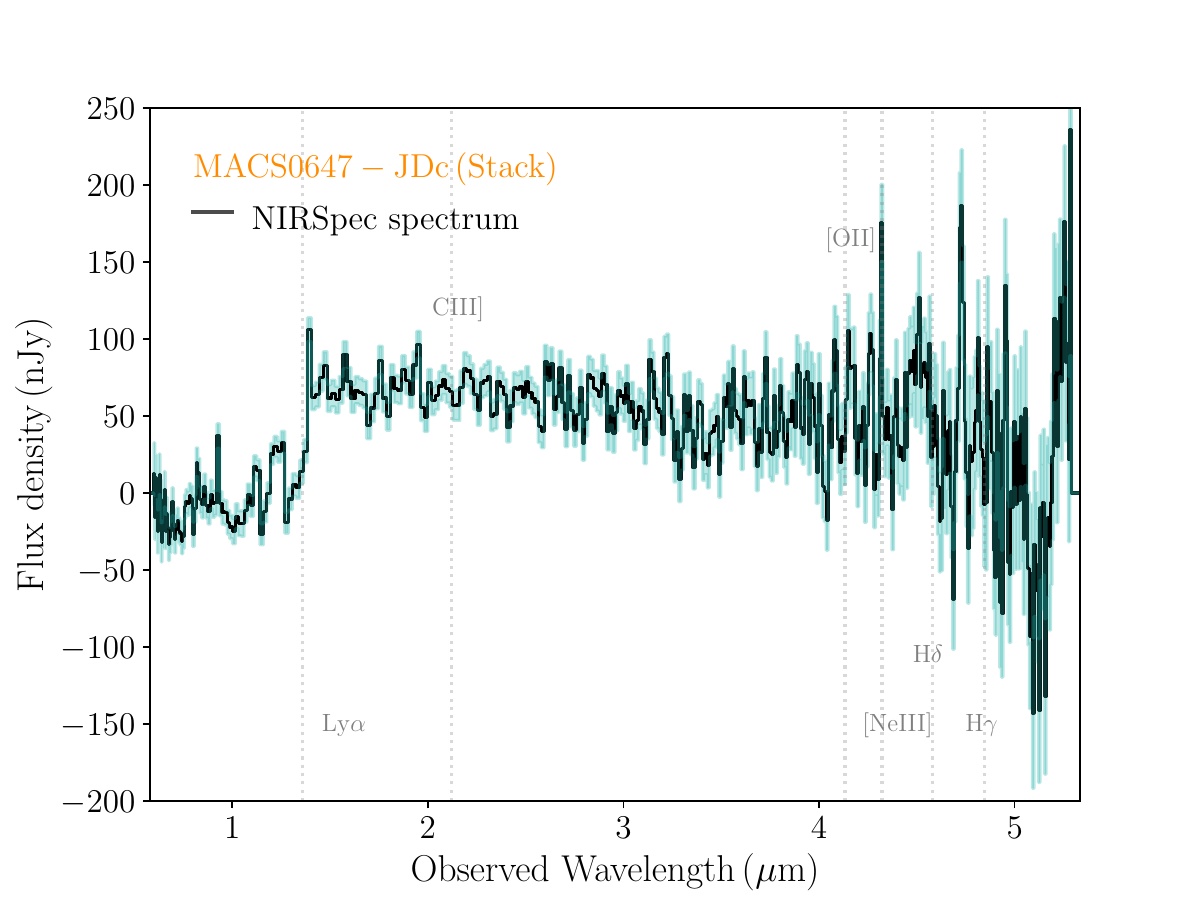}
\caption{{sum} of the spectra shown in Figure \ref{fig:JDC_spectra} 
for two lensed images of the candidate companion galaxy C.
No emission lines are confidently detected;
their expected wavelengths are labeled 
assuming a redshift $z = 10.17$ as derived for \JD.
\label{fig:companion}
}
\end{figure*}

\subsection{Candidate companion galaxy C}
\label{sec:companion}
As discovered in \citet{Hsiao2023}, 
a nearby triply-lensed J-band dropout galaxy
is suspected to be a companion of \JD\ \about 3$\,$kpc away.
The photometric redshift and geometric lens model redshift 
of this companion ``C'' are consistent with \JD\ {of 10.17};
the observed lensed positions require a similar or identical redshift.
Here we show the NIRSpec spectra of the candidate companion C
in Figure \ref{fig:JDC_spectra}
and the 1D {sum} in Figure \ref{fig:companion}.
The Obs 21 spectrum shows a clear Lyman break consistent with $z = 10.17$.
The Obs 23 spectrum is noisier but also consistent with the redshift of MACS0647--JD ($z=10.17$).
This supports our conclusion from \cite{Hsiao2023}
that \JD\ is likely the most distant galaxy merger known at $z = 10.17$,
including two merging components A and B plus a third component C nearby, where component C may indeed be a companion of A and B components.


Given the lower SNR spectrum of the companion C,
we do not detect emission lines, nor do we detect the {\Lya} damping wing
expected due to the proximity to A.



\section{Conclusions}
\label{sec:conclusions}
In this study, we report NIRSpec prism observations of MACS0647$–$JD,
primarily targeting its small component A with delensed radius $r \sim 70\,$pc. 
We measure a spectroscopic redshift $z=10.17$
based on detections of 7 emission lines:
\CIIIdw, \OIIw, \NeIIIw, \NeIIIwb, \Hdelta, \Hgamma, and \OIIIwa.

This redshift is slightly lower than the photometric redshifts $z \sim 10.6$
estimated previously based on \HST\ and \JWST+\HST\ imaging.
This may be explained by a \Lya\ damping wing revealed by NIRSpec
that suppresses the photometry at \about 1.5$\,$\um\ more than expected for a sharp Lyman break.
Based on observed and extrapolated line ratios of JDA,
we estimate an ionization parameter \logU\ $= -1.9 \pm 0.2$,
and ionizing photon production efficiency log($\xi_{\rm ion})=25.2\pm0.2$ for clump A.
This \logxiion\ value is similar to values measured in sub-\Lstar\ $z \sim 4$ -- 5 galaxies
\citep{Bouwens2016}
but lower than measured in $z \sim 7$ -- 8 galaxies:
\logxiion\ = 25.7 -- 26.0 \citep{Tang2023}.



Based on the observed and extrapolated line ratios of JDA,
we also estimate a gas phase metallicity \logOH\ = 7.5 -- 8.0, or (0.06 -- 0.2) \Zsun\, 
derived using 6 different methods,
including optical line diagnostics
as well as one diagnostic based on the rest-UV line \CIIId\ \citep{Mingozzi2022}.
Derived metallicities are consistent across different methods.
The overall agreement is encouraging, 
supporting the use of each individual method in galaxies with fewer detected emission lines.
Additional MIRI spectroscopy covering \OIIIww, \Hbeta, and \Halpha\ would enable
a more precise ``direct'' metallicity measurement
when combined with the other lines presented here, including the auroral line \OIIIwa.

The total stellar mass of \JD\ is log($M / M_*$) $= 8.1 \pm 0.3$
based on SED fitting to the photometry of the entire galaxy including both clumps A and B.
This is towards the low end of stellar masses for $z = 4$ -- 10 galaxies
with direct metallicity estimates from \JWST\ NIRSpec spectroscopy \citep{Nakajima2023}.
The metallicity of \JD\ at $z = 10.17$
(if we assume the metallicity of clump A is representative of the whole galaxy)
is consistent with the metallicities measured for $4 < z < 10$ galaxies of similar stellar mass.



\JD\ has a delensed absolute UV magnitude $M_{UV} = -20.3 \pm 0.2$,
including $M_{UV} = -19.5 \pm 0.2$ for clump A
and $M_{UV} = -17.9 \pm 0.2$ for clump B.
SED fitting to the observed photometry of clump A yields a 
stellar mass log($M / M_*$) $= 7.8 \pm 0.3$,
likely all produced within the most recent 20 Myr.
This young age estimate is supported by the 
observed strong \CIIIdw\ with rest-frame equivalent width $14 \pm 2$ \AA.
The star formation rate of clump A based on \Hgamma\ and assuming no dust
is $1.4\pm0.2\,M_{\odot}/{\rm yr}$ over the past 10 Myr.
This is similar to the estimate $2\pm1\,M_{\odot}/{\rm yr}$ 
based on SED fitting to clump A's photometry.

%

Clump B is likely older with a rising star formation history
over \about 50 Myr or more,
stellar mass log($M/M_*$) = $7.8 \pm 0.3$,
and $0.13 \pm 0.07$ mag dust.
The older age of clump B suggests it may have merged with clump A
rather than the two clumps forming together in the same galaxy \citep{Hsiao2023}.
Spectroscopy of a nearby galaxy C \about 3 kpc away (delensed)
reveals a Lyman break consistent with \JD's redshift $z = 10.17$.
This companion C may be destined to merge with A and B.
\JD\ is likely the most distant galaxy merger yet known.




Emission line spectroscopy is possible for the triply-lensed galaxy \JD\ thanks to 
its 3 images magnified by factors \about 8, 5, and 2.
The brightest lensed image JD1 is F200W AB mag {$25.0\pm0.1$},
including AB mag {$25.8\pm0.1$} for clump A.
Of all known $z > 10$ galaxies, 
only GN-z11 at $z = 10.603$ \citep{Bunker2023}
is similarly bright ({F200W} AB mag {$26.0\pm0.1$}; \citealt{Tacchella2023}),
while most others are AB mag 28 -- 29 \citep{CurtisLake2023,Arrabal-Haro2023a}.
\JD\ and GN-z11 offer unique opportunities for detailed studies 
of galaxies within the first 500 Myr.

Many of the physical properties presented in this work
rely on extrapolations for \OIIIw, \Hbeta\, and/or \Halpha\ line fluxes.
{With our granted JWST time in cycle 2, we will obtain MIRI MRS spectroscopy covering $\sim 5-7.7\micron$ of observed wavelength (rest-frame of $0.45-0.7 \micron$) which will reveal those emission lines (though \Hbeta\ is likely too faint for detection) and enable a more robust analysis of this galaxy. The observation will give spatially resolved spectra that cover both clumps A and B as well as component C of the JD1 system, which gives an opportunity to further confirm the merging nature of this system. In addition to this, we will also obtain high-resolution NIRSpec spectroscopy to better resolve the emission lines currently detected in our prism spectrum, enabling a better measurement of the equivalent width.}  





\section*{Acknowledgments}

{We thank the anonymous referee for useful comments and constructive remarks on the manuscript.}

We are grateful and indebted to the 20,000 people who worked to make \JWST\ an incredible discovery machine.

This work is based on observations made with the NASA/ESA/CSA 
\textit{James Webb Space Telescope} (\JWST)
and \textit{Hubble Space Telescope} (\HST). 
The data were obtained from the Mikulski Archive for Space Telescopes (MAST) 
at the Space Telescope Science Institute (STScI), 
which is operated by the Association of Universities for Research in Astronomy (AURA), Inc., 
under NASA contract NAS 5-03127 for \JWST. 

These observations are associated with programs
\JWST\ GO 1433 and \HST\ GO 9722, 10493, 10793, and 12101.

TH and A were funded by a grant for JWST-GO-01433 provided by STScI under NASA contract NAS5-03127.
AA acknowledges support from the Swedish Research Council (Vetenskapsr\aa{}det project grants 2021-05559).
PD acknowledges support from the NWO grant 016.VIDI.189.162 (``ODIN") and the European Commission's and University of Groningen's CO-FUND Rosalind Franklin program.  The Cosmic
Dawn Center is funded by the Danish National Research Foundation (DNRF) under grant \#140. 
EZ and AV acknowledge support from the Swedish National Space Agency. EZ also acknowledges grant 2022-03804 from the Swedish Research Council.
MB acknowledges support from the Slovenian national research agency ARRS through grant N1-0238.
TAH is supported by an appointment to the NASA Postdoctoral Program (NPP) at NASA Goddard Space Flight Center, administered by Oak Ridge Associated Universities under contract with NASA.
AZ and LJF acknowledge support by Grant No. 2020750 from the United States-Israel Binational Science Foundation (BSF) and Grant No. 2109066 from the United States National Science Foundation (NSF), and by the Ministry of Science \& Technology, Israel.
EV acknowledge financial support through grants PRIN-MIUR 2017WSCC32, 2020SKSTHZ and INAF “main-stream” grants 1.05.01.86.20 and 1.05.01.86.31. EV acknowledges support from
the INAF GO Grant 2022 “The revolution is around the corner: JWST will probe globular cluster precursors and Population III stellar clusters at cosmic dawn”.
ACC thanks the Leverhulme Trust for their support via a Leverhulme Early Career Fellowship.


\facilities{JWST(NIRCam, NIRSpec), HST(ACS, WFC3)}


\software{STScI JWST pipeline;
          \msaexp;
          \grizli\ \citep{grizli};
          \eazypy\ \citep{Brammer2008};
          \astropy\ \citep{astropy2022, astropy2018, astropy2013};
          \photutils\ \citep{Bradley2022};
          \piXedfit\ \citep{Abdurrouf2021,Abdurrouf2022};
          \bagpipes\ \citep{Carnall2018, Carnall2019b};
          \beagle\ \citep{Chevallard2016_BEAGLE};
          \cloudy\ \citep{Ferland1998,Ferland2013,Ferland2017};
          \pyneb\ \citep{pyneb_Luridiana2015};
          \HIIC\ \citep{Perez2014}
          }



\bibliography{papers}{}
\bibliographystyle{aasjournal}


\end{document}

%% file: newcommands.tex
\newcommand{\LCDM}{$\Lambda$CDM}

\newcommand{\red}[1]{{\color{red} #1}}
\newcommand{\redss}[1]{{\color{red} ** #1}}
\newcommand{\redbf}[1]{{\color{red}\bf #1 \color{black}}}

\newcommand{\ny}{$\tilde {\rm n}$}
\newcommand{\about}{$\sim$}
\newcommand{\appr}{$\approx$}
\newcommand{\gt}{$>$}
\newcommand{\um}{$\mu$m}
\newcommand{\uJy}{$\mu$Jy}
\newcommand{\sig}{$\sigma$}
\newcommand{\Lya}{Lyman-$\alpha$}
\renewcommand{\th}{$^{\rm th}$}
\newcommand{\lam}{$\lambda$}

\newcommand{\tentothe}[1]{$10^{#1}$}
\newcommand{\tentotheminus}[1]{$10^{-#1}$}
\newcommand{\e}[1]{$\times 10^{#1}$}
\newcommand{\en}[1]{$\times 10^{-#1}$}
\newcommand{\cgsfluxunits}{erg$\,$s$^{-1}\,$cm$^{-2}$}
\newcommand{\linefluxunits}{\tentotheminus{20} \cgsfluxunits}

\newcommand{\logU}{$\log(U)$}
\newcommand{\logOH}{12+log(O/H)}

\newcommand{\sinv}{s$^{-1}$}

\newcommand{\footnoteurl}[1]{\footnote{\url{#1}}}

\newcommand{\tnm}[1]{\tablenotemark{#1}}
\newcommand{\super}[1]{$^{\rm #1}$}
\newcommand{\supa}{$^{\rm a}$}
\newcommand{\supb}{$^{\rm b}$}
\newcommand{\supc}{$^{\rm c}$}
\newcommand{\supd}{$^{\rm d}$}
\newcommand{\supe}{$^{\rm e}$}
\newcommand{\supf}{$^{\rm f}$}
\newcommand{\supg}{$^{\rm g}$}
\newcommand{\suph}{$^{\rm h}$}
\newcommand{\supi}{$^{\rm i}$}
\newcommand{\supj}{$^{\rm j}$}
\newcommand{\supk}{$^{\rm k}$}
\newcommand{\supl}{$^{\rm l}$}
\newcommand{\supm}{$^{\rm m}$}
\newcommand{\supn}{$^{\rm n}$}
\newcommand{\supo}{$^{\rm o}$}

\newcommand{\squared}{$^2$}
\newcommand{\cubed}{$^3$}

\newcommand{\sqarcmin}{arcmin\squared}

\newcommand{\supcomma}{$^{\rm ,}$}

\newcommand{\rhalf}{$r_{1/2}$}

\newcommand{\chisq}{$\chi^2$}

\newcommand{\Zgas}{$Z_{\rm gas}$}  
\newcommand{\Zstar}{$Z_*$}  

\newcommand{\per}{$^{-1}$}
\newcommand{\inv}{\per}
\newcommand{\Mstar}{$M^*$}
\newcommand{\Lstar}{$L^*$}
\newcommand{\phistar}{$\phi^*$}

\newcommand{\logM}{log($M_*$/\Msun)}

\newcommand{\LUV}{$L_{UV}$}
\newcommand{\MUV}{$M_{UV}$}

\newcommand{\Msun}{$M_\odot$}
\newcommand{\Lsun}{$L_\odot$}
\newcommand{\Zsun}{$Z_\odot$}

\newcommand{\Mvir}{$M_{vir}$}
\newcommand{\Mt}{$M_{200}$}
\newcommand{\Mf}{$M_{500}$}

\newcommand{\Ndotion}{$\dot{N}_{\rm ion}$}
\newcommand{\xiion}{$\xi_{\rm ion}$}
\newcommand{\logxiion}{log(\xiion)}
\newcommand{\fesc}{$f_{\rm esc}$}

\newcommand{\XHI}{$X_{\rm HI}$}
\newcommand{\XHII}{$X_{\rm HII}$}
\newcommand{\RHII}{$R_{\rm HII}$}

\newcommand{\Halpha}{H$\alpha$}
\newcommand{\Hbeta}{H$\beta$}
\newcommand{\Hgamma}{H$\gamma$}
\newcommand{\Hdelta}{H$\delta$}
\newcommand{\Halphaw}{\Halpha\,$\lambda$6563}
\newcommand{\Hbetaw}{\Hbeta\,$\lambda$4861}
\newcommand{\Hgammaw}{H$\gamma$\,$\lambda$4340}
\newcommand{\Hdeltaw}{H$\delta$\,$\lambda$4101}
\newcommand{\Ha}{\Halpha}
\newcommand{\Hb}{\Hbeta}

\newcommand{\I}{\,{\sc i}}
\newcommand{\II}{\,{\sc ii}}
\newcommand{\III}{\,{\sc iii}}
\newcommand{\IV}{\,{\sc iv}}
\newcommand{\V}{\,{\sc v}}
\newcommand{\VI}{\,{\sc vi}}
\newcommand{\VII}{\,{\sc vii}}
\newcommand{\VIII}{\,{\sc viii}}

\newcommand{\HI}{H\,{\sc i}}
\newcommand{\HII}{H\,{\sc ii}}
\newcommand{\HeI}{He\,{\sc i}}
\newcommand{\HeII}{He\,{\sc ii}}

\newcommand{\CII}{[C\,{\sc ii}]}
\newcommand{\CIIw}{\CII\,$\lambda$2325 (blend)}
\newcommand{\CIII}{[C\,{\sc iii}]}
\newcommand{\CIIIw}{\CIII\,$\lambda$1908}
\newcommand{\CIIId}{C\,{\sc iii}]}
\newcommand{\CIIIdw}{C\,{\sc iii}]\,$\lambda\lambda$1907,1909}
\newcommand{\CIV}{C\,{\sc iv}}
\newcommand{\CIVw}{\CIV\,$\lambda$1549}
\newcommand{\OII}{[O\,{\sc ii}]}
\newcommand{\OIIw}{\OII\,$\lambda$3727}
\newcommand{\OIIdw}{\OII\,$\lambda\lambda$3727,3729}
\newcommand{\OIII}{[O\,{\sc iii}]}
\newcommand{\OIIIw}{\OIII\,$\lambda$5007}
\newcommand{\OIIIww}{\OIII\,$\lambda$4959,$\lambda$5007}
\newcommand{\OIIIwa}{\OIII\,$\lambda$4363}
\newcommand{\OIIIwc}{\OIII\,$\lambda$4959}
\newcommand{\NeIII}{[Ne\,{\sc iii}]}
\newcommand{\NeIIIw}{\NeIII\,$\lambda$3869}
\newcommand{\NeIIIwb}{\NeIII\,$\lambda$3968}
\newcommand{\HeIw}{HeI\,$\lambda$3889}
\newcommand{\HeIwa}{HeI\,$\lambda$4473}
\newcommand{\HeIIw}{HeII\,$\lambda$1640}
\newcommand{\NII}{[N\,{\sc ii}]}
\newcommand{\NIII}{N\,{\sc iii}]}
\newcommand{\NIV}{N\,{\sc iv}]}
\newcommand{\NIIIw}{\NIII\,$\lambda$1748}
\newcommand{\NIVw}{\NIV\,$\lambda$1486}
\newcommand{\MgII}{Mg\,{\sc ii}}
\newcommand{\MgIIw}{\MgII\,$\lambda$2800}

\newcommand{\Lyaw}{Ly$\alpha$\,$\lambda$1216}



\newcommand{\Om}{\Omega_{\rm M}}
\newcommand{\OL}{\Omega_\Lambda}

\newcommand{\etal}{et al.}

\newcommand{\citeps}{\citep}

\newcommand{\HST}{{\em HST}}
\newcommand{\SST}{{\em SST}}
\newcommand{\Hubble}{{\em Hubble}}
\newcommand{\Spitzer}{{\em Spitzer}}
\newcommand{\Chandra}{{\em Chandra}}
\newcommand{\JWST}{{\em JWST}}
\newcommand{\Planck}{{\em Planck}}

\newcommand{\Bradac}{{Brada\v{c}}}

\newcommand{\citepeg}[1]{\citep[e.g.,][]{#1}}

\newcommand{\range}[2]{\! \left[ _{#1} ^{#2} \right] \!}  

\newcommand{\grizli}{\textsc{grizli}}
\newcommand{\eazypy}{\textsc{eazypy}}
\newcommand{\msaexp}{\textsc{msaexp}}
\newcommand{\trilogy}{\textsc{trilogy}}
\newcommand{\bagpipes}{\textsc{bagpipes}}
\newcommand{\beagle}{\textsc{beagle}}
\newcommand{\photutils}{\textsc{photutils}}
\newcommand{\SEP}{\textsc{sep}}
\newcommand{\piXedfit}{\textsc{piXedfit}}
\newcommand{\pyneb}{\textsc{pyneb}}
\newcommand{\HIIC}{\textsc{hii-chi-mistry}}
\newcommand{\astropy}{\textsc{astropy}}
\newcommand{\astrodrizzle}{\textsc{astrodrizzle}}
\newcommand{\multinest}{\textsc{multinest}}
\newcommand{\cloudy}{\textsc{Cloudy}}
\newcommand{\jdaviz}{\textsc{Jdaviz}}

\renewcommand{\tt}[1]{\texttt{#1}}

\newcommand{\SE}{\tt{SourceExtractor}}

\newcommand{\PD}[1]{\textcolor{blue}{[PD: #1\;]}}

\newcommand{\JD}{MACS0647$-$JD}

%% file: observations.tex
\begin{deluxetable*}{lccccc}
\tablecaption{\label{tab:obs}\JWST\ Observations of MACS0647}
\tablewidth{\columnwidth}
\tablehead{
\colhead{Observation}&
\colhead{Date (UT)}&
\colhead{Mode}&
\colhead{Exposures}&
\colhead{Filters $/$ Element}&
\colhead{Exposure Time\supa}
}
\startdata
10 & 2022-09-23 & NIRCam imaging & 3 filter pairs & F115W, F150W, F200W, & 2104 s\\
 & & &  & F277W, F356W, F444W &  \\
20 & 2023-01-08 & NIRCam imaging & 1 filter pair & F200W, F480M & 2104 s\\
21 & 2023-01-08 & NIRSpec MOS & 1 slitlet; 2 dithers\supb & prism & 6420 s\\
23 & 2023-02-20 & NIRSpec MOS & 3 slitlet nods & prism & 6567 s\\
\enddata
\tablenotetext{a}{Exposure time per imaging filter; 
total exposure times for NIRSpec}
\tablenotetext{b}{Each dither requires its own MSA configuration.}
\end{deluxetable*}

%% file: photometry.tex
\begin{deluxetable*}{lcllccccccc}
\tablecaption{\label{tab:pho}NIRCam photometry of \JD}
\tablewidth{\columnwidth}
\tablehead{
\colhead{Object}&
\colhead{Magnif.\supa}&
\colhead{Photometry}&
\colhead{Aperture}&
\colhead{F115W}&
\colhead{F150W}&
\colhead{F200W}&
\colhead{F277W}&
\colhead{F356W}&
\colhead{F444W}&
\colhead{F480M}
}
\startdata
JD1 & \about 8 & \grizli & circular & $1\pm8$ &  $297\pm7$ & $368\pm5$ & $309\pm6$ & $306\pm6$ & $317\pm8$ &  $314\pm18$ \\
JD2 & \about 5.3 & \grizli & circular & $12\pm7$ &  $178\pm6$ & $255\pm4$ & $229\pm5$ & $205\pm5$ & $209\pm7$ & $185\pm16$ \\
JD1A & \about 8 & \piXedfit & elliptical & $-8\pm4$ & $146\pm3$& $169\pm2$ & $135\pm2$ & $120\pm2$ & $116\pm3$ & $124\pm8$ \\
JD2A & \about 5.3 & \piXedfit & elliptical & $-2\pm3$ & $90\pm3$ & $125\pm2$ & $95\pm2$ & $82\pm2$ & $77\pm3$ & $87\pm7$ \\
JD1B & \about 8 & \piXedfit & elliptical & $1\pm2$ & $26\pm2$& $40\pm2$ & $31\pm2$ & $39\pm2$ & $44\pm2$ & $45\pm5$ \\
JD2B & \about 5.3 & \piXedfit & elliptical & $-3\pm$3 & $17\pm2$& $31\pm1$ & $27\pm1$ & $27\pm1$ & $28\pm2$ & $29\pm5$ \\
JDA\supb & \about 26.6 & \piXedfit & rectangular & $-35\pm10$ &  $450\pm8$ & $623\pm5$ & $509\pm6$ & $461\pm6$ & $441\pm9$ & $446\pm21$\\
 \enddata
\tablenotetext{a}{Lensing magnification estimates \citep{Meena2023}}
\tablenotetext{b}{Photometric apertures designed to match the 4 observed NIRSpec slitlets that primarily observe JDA. 
Flux densities include the sum of all 4.}
\tablenotetext{}{{\bf Note.} All flux densities are given in $\rm nJy$, {and are not de-lensed}.
Photometry is plotted in Figure \ref{fig:spectrum-photometry} along with the spectrum.
}
\end{deluxetable*}

%% file: line_flux.tex
\begin{deluxetable*}{lccrcc}
\tablecaption{\label{tab:lines}Measured emission line fluxes for JDA
}
\tablewidth{\columnwidth}
\tablehead{
\colhead{Emission Lines} &
\colhead{Rest wavelength} &
\colhead{Observed wavelength} &
\colhead{Line fluxes\supa\ } &
\colhead{SNR} &
\colhead{Equivalent width}
\vspace{-0.07in}\\
\colhead{} &
\colhead{$\rm \AA$}  &
\colhead{$\rm \mu m$} &
\colhead{$10^{-20}\,$erg/s/cm$^2$} &
\colhead{} &
\colhead{(rest-frame) $\rm \AA$} 
}
\startdata
\CIIId
& 1907, 1909
& 2.13
& $310\pm60$
& $5.7$
&  14$\pm3$\\
\OII
& 3726, 3729
& 4.16
& $42\pm5$ 
& $8.9$ 
&  13$\pm2$\\
\NeIII
& 3869 
& 4.32 
& $76\pm6$
& $13.8$
&  25$\pm2$\\
\NeIII
& 3968
& 4.43
& $41\pm6$
& $7.3$
& 15$\pm2$\\
\Hdelta
& 4101 
& 4.58 
& $40\pm7$
& $6.1$ 
&  17$\pm3$\\
\Hgamma
& 4340
& 4.85
& $74\pm10$
& $7.7$
& 42$\pm6$\\
\OIII
& 4363 
& 4.87
& $25\pm6$ 
& $4.4$ 
&  14$\pm3$\\
HeI
& 4473 
& 5.00
& $24\pm9$ 
& $2.8$ 
& 16$\pm5$
\enddata
\tablenotetext{a}{As observed in the {sum} of 4 observations {($=2\mu_{\rm JD1}+2\mu_{\rm JD2} = 2\times (8 + 5.3)$)} divided by 2 with a total magnification \about 13.3 {($=\mu_{\rm JD1}+\mu_{\rm JD2} = 8 + 5.3$), which is not de-lensed.}}
\end{deluxetable*}

%% file: individual.tex
\begin{deluxetable*}{llcccccc}
\tablecaption{\label{tab:indi_phy}Physical properties of JD1, JD1A, JD1B from SED fitting using various methods}
\tablewidth{\columnwidth}
\tablehead{
\colhead{} &
\colhead{} &
\colhead{} &
\colhead{Age$^a$} &
\colhead{Stellar Mass} &
\colhead{SFR} &
\colhead{specific SFR} &
\colhead{Dust}
\vspace{-0.07in}\\
\colhead{Component} &
\colhead{Code} &
\colhead{Star Formation History} &
\colhead{Myr} &
\colhead{log($M_{*}/M_\odot$)} &
\colhead{$M_{*}$/yr} &
\colhead{Gyr\inv} &
\colhead{$A_{V}$ mag} 
}
\startdata
JD1   & \bagpipes  & constant & $8^{+3}_{-2}$ & $8.1 \pm 0.3$ & $10 \pm 3$ & $-7.1 \pm 0.3$ & $0.09 \pm 0.02$\\
      & \bagpipes  & non-parametric & $6^{+12}_{-1}$ & $8.0\pm0.1$ & $10 \pm 1$ & $-7.0 \pm 0.1$ & $0.09 \pm 0.01$\\
      & \bagpipes  & delayed $\tau$ & $3^{+3}_{-2}$ & $7.7^{+0.2}_{-0.1}$ & $5.5^{+0.8}_{-1.0}$ & $-7.0\pm0.1$ & $0.02^{+0.02}_{-0.01}$\\
      & \beagle     & delayed $\tau$ & $40^{+85}_{-25}$ & $8.2\pm0.1$ & $10\pm1$ & $-7.2\pm0.1$ &$0.11\pm0.03$\\
      & \piXedfit  & double power-law & $30^{+53}_{-20}$ & $8.4\pm0.3$ & $4.4^{+2.5}_{-1.2}$ & $-7.7^{+0.1}_{-0.2}$ & $0.05\pm0.04$\\
\hline  
JD1A & \bagpipes   &   constant & $8^{+3}_{-2}$ & $7.6\pm0.1$ & $3 \pm 1$ & $-7.0 \pm 0.3$ & $<0.01$ \\
     & \bagpipes   & delayed $\tau$ & $5^{+2}_{-1}$ & $7.5\pm0.1$ & $2.4\pm0.1$ & $-7.0\pm0.1$ & $<0.01$ \\
     & \beagle     & delayed $\tau$ & $14^{+6}_{-3}$ & $7.9\pm0.1$ & $3.3\pm0.1$ & $-7.3\pm0.1$ & $0.05\pm0.01$\\
     & \piXedfit   & double power-law & $38^{+28}_{-20}$  & $8.1\pm0.3$ & $1.3\pm0.6$ & $-8.0\pm0.1$ & $0.03^{+0.04}_{-0.03}$\\
\hline  
JD1B & \bagpipes   & delayed $\tau$   & $50^{+60}_{-40}$ & $7.7\pm0.3$ & $1.0\pm0.3$ & $-7.7\pm0.3$ & $0.13 \pm 0.07$ \\
     & \piXedfit   & double power-law & $15^{+40}_{-10}$ & $7.3\pm0.3$ & $0.7^{+0.5}_{-0.3}$ & $-7.4^{+0.1}_{-0.2}$ & $0.06\pm0.04$\\
\enddata
\tablenotetext{a}{Ages are mass-weighted. This equals half the formation age for a constant star formation history.}
\textbf{Note.} 
Stellar mass and SFR estimates are delensed by a magnification $\mu = 8 \pm 1$ \citep{Meena2023}.
Magnification uncertainties are not included in the uncertainties quoted in the table for stellar mass and SFR.
Most of these SED fitting methods assume a \citet{Chabrier2003} IMF.
\bagpipes\ assumes \citet{Kroupa1993};
to renormalize those results, we multiplied the stellar masses by 0.94
\citep{Madau2014}. {Component JD1 is measured with \grizli\ while JD1A and JD1B are measured using \piXedfit\ (see Table \ref{tab:pho}). The summed SFR of JD1A and JD1B is $\sim$ an order of magnitude lower than the combined JD1, which is likely due to lost fluxes among various apertures (see also Table \ref{tab:pho}).}

\end{deluxetable*}

%% file: line_ratios.tex
\begin{deluxetable*}{lcc}
\tablecaption{\label{tab:ratio}Derived emission line ratios and expected values
}
\tablewidth{\columnwidth}
\tablehead{
\colhead{Line or line ratio} &
\colhead{Equation / Derivation} &
\colhead{Flux or flux ratio\supa}
}
\startdata
\Hbeta  & $2.14$ \Hgamma            & ($160  \pm 20$) $\times$ \linefluxunits    \\
\Halpha & $6.11$ \Hgamma            & ($460  \pm 60$) $\times$ \linefluxunits   \\
Ne3O2   & \NeIIIw\ $/$ \OIIw         & $1.8 \pm 0.2$  \\
O32        & from Ne3O2\supb; \OIIIw\ $/$ \OII     & $30 \pm 6$ \\
\OIIIw  & O32 $\times$ \OII        & ($1300 \pm 300$) $\times$ \linefluxunits \\
auroral line ratio  &   \OIIIw $/$ \OIIIwa  & $50 \pm 16$\\
\OIIIww  &  1.34 $\times$ \OIIIw       & ($1700 \pm 400$) $\times$ \linefluxunits \\
R23     & (\OII\ + \OIIIww) $/$ \Hbeta & $11 \pm3$ \\
\enddata
\tablenotetext{a}{Measured fluxes in the stacked spectra are magnified by \about 13.3}
\tablenotetext{b}{We estimate O32 based on our measured Ne3O2. 
The two line ratios are strongly correlated at these high values.}
\end{deluxetable*}

%% file: metallicites.tex
\begin{deluxetable*}{llcc}
\tablecaption{\label{tab:metal}Metallicity Estimates}
\tablewidth{\columnwidth}
\tablehead{
\colhead{Method}&
\colhead{Emission Lines}&
\colhead{\logOH}&
\colhead{$Z$}
}
\startdata
\pyneb       &  \OII, \OIIIwa;          extrapolated${^a}$ \Hbeta, \OIIIw       & $8.0 \pm 0.2$  &  $0.20^{+0.12}_{-0.07}$ \Zsun \\
\HIIC        &  \OII, \NeIIIw, \OIIIwa; extrapolated${^a}$ \Hbeta               & $7.5\pm0.2$    &  $0.06^{+0.04}_{-0.02}$ \Zsun \\
Ne3O2        &  \OII, \NeIIIw                                                & $7.6 \pm 0.3$  &  $0.08^{+0.08}_{-0.04}$ \Zsun \\
R23          &  \OII;                   extrapolated${^a}$ \Hbeta, \OIIIw       & $8.0 \pm 0.3$  &  $0.20^{+0.20}_{-0.10}$ \Zsun \\
R23 and O32  &  \OII;                   extrapolated${^a}$ \Hbeta, \OIIIw       & $7.7\pm0.2$    &  $0.10^{+0.06}_{-0.04}$ \Zsun \\
EW(\CIIId)   &  \CIIId                                                       & $7.8\pm0.2$    &  $0.13^{+0.07}_{-0.05}$ \Zsun
\enddata
\tablenotetext{a}{Extrapolated line fluxes correspond to those derived from either the theoretical line ratios (e.g., H$\beta$/H$\gamma$) or empirical calibrations (e.g., Ne3O2 vs O3O2). }
\end{deluxetable*}

%% file: properties.tex
\begin{deluxetable*}{lcccc}
\tablecaption{\label{tab:phy}Physical properties of \JD\ (entire galaxy)
and individual clumps A and B,
estimated based on observed and estimated line fluxes and SED fitting to the photometry.
Star formation rate (average within the past 10 Myr) 
and stellar mass are corrected for lensing magnification.}
\tablewidth{\columnwidth}
\tablehead{
\colhead{Parameter}&
\colhead{Variable}&
\colhead{MACS0647-JD} &
\colhead{clump A} &
\colhead{clump B}
}
\startdata
\multicolumn{5}{c}{Emission lines} \\ \hline
\multirow{2}{*}{Metallicity (gas phase)} & \logOH & & 7.5 -- 8.0 & \\
 & $Z_{\rm gas} / Z_\odot$ & &  0.06 -- 0.2 &\\
Ionization parameter & \logU && $-1.9\pm0.2$& \\
Ionizing photon production efficiency&log($\xi_{\rm ion} /$erg\inv\,Hz)& & $25.2 \pm 0.2$ & \\
Star formation rate & SFR (${\rm M_{\odot}/yr}$) & &$1.4\pm0.2$&\\ 
\hline \hline  
\multicolumn{5}{c}{SED fitting} \\ 
\hline  
Absolute UV magnitude & $M_{UV}$ (AB mag) & $-20.3\pm0.2$ & $-19.5\pm0.2$ & $-17.9\pm0.2$ \\
Star formation rate & SFR (${\rm M_{\odot}/yr}$) & $8\pm3$ & $2\pm1$ & $0.9\pm0.3$\\
Stellar mass & log($M_* / M_\odot$) & $8.1 \pm 0.3$  &$7.8 \pm 0.3$ & $7.5\pm0.3$\\
Specific star formation rate & log(sSFR/Gyr\inv) & $-7.1\pm0.3$ & $-7.2\pm0.3$ & $-7.5\pm0.3$ \\
Mass-weighted age & ${\rm age}_{\rm M}\,$(Myr) & $20 \pm 20$   & $10^{+30}_{-9}$ & $40 \pm 30$ \\
Dust & $A_V$ (mag) & $0.07\pm0.06$ & $<$ 0.06 & $ 0.10\pm0.05$\\
 \enddata
\tablenotetext{}{
{{\textbf{Note}. BAGPIPES assumes \citet{Kroupa1993};
to renormalize those results, we multiplied the stellar masses by 0.94
\citep{Madau2014}. 
}}}
\end{deluxetable*}